\documentclass[traditabstract]{aa} 
\usepackage[utf8]{inputenc}
\usepackage{graphicx}
\pdfoutput=1

\usepackage{amsmath}
\usepackage{amsfonts}
\usepackage{txfonts}
\usepackage{epstopdf}
\usepackage{times,epsfig}
\usepackage{mathrsfs}
\usepackage{natbib}
\usepackage{amssymb}
\bibpunct{(}{)}{;}{a}{}{,} 
\usepackage[english]{babel}
\usepackage{longtable}
\usepackage[dvipsnames]{xcolor}
\usepackage[]{hyperref}
\usepackage{deluxetable}
\usepackage{float}
\usepackage{subcaption}
\usepackage[switch]{lineno}

\colorlet{myurlcolor}{violet}
\colorlet{myallcolor}{MidnightBlue}
\hypersetup{
  linkcolor  = myallcolor,
  citecolor  = myallcolor,
  urlcolor   = myurlcolor,
  colorlinks = true,
}
\usepackage{graphicx}
\usepackage{txfonts}

\usepackage{amssymb}
\usepackage[T1]{fontenc}
\usepackage{textcomp}
\usepackage{longtable}
\usepackage{booktabs}
\usepackage{xcolor}
\usepackage{lscape}
\usepackage[switch]{lineno}
\usepackage{orcidlink}
\DeclareUnicodeCharacter{2212}{-}

\usepackage{natbib}
\newcommand{\msun}{M$_{\sun}$}
\usepackage{xspace}
\newcommand\sn{SN~2020aeuh\xspace}

\usepackage{hyperref}

\begin{document}

\title{A thermonuclear supernova interacting with hydrogen- and helium-deficient circumstellar material}
\subtitle{\sn as a SN Ia-CSM-C/O?}

\author{K. Tsalapatas\inst{1}\orcidlink{0009-0004-1062-8886}\thanks{{email: konstantinos.tsalapatas@astro.su.se}}
\and{J. Sollerman}\inst{1}\orcidlink{0000-0003-1546-6615}
\and{R. Chiba}\inst{2,3}\orcidlink{0009-0003-4594-3715}
\and{E. Kool}\inst{1}
\and{J. Johansson}\inst{4}\orcidlink{0000-0001-5975-290X}
\and{S. Rosswog}\inst{1,5}\orcidlink{0000-0002-3833-8520}
\and{S. Schulze}\inst{6}\orcidlink{0000-0001-6797-1889}
\and{T. J. Moriya}\inst{2,3,7}\orcidlink{0000-0003-1169-1954}
\and{I. Andreoni}\inst{8}\orcidlink{0000-0002-8977-1498}
\and{T. G. Brink}\inst{9}\orcidlink{0000-0001-5955-2502}
\and{T. X. Chen}\inst{10}\orcidlink{0000-0001-9152-6224}
\and{S. Covarrubias}\inst{11}\orcidlink{0000-0003-1858-561X}
\and{K. De}\inst{12}\orcidlink{0000-0002-8989-0542}
\and{G. Dimitriadis}\inst{13}\orcidlink{0000-0001-9494-179X}
\and{A. V. Filippenko}\inst{9}\orcidlink{0000-0003-3460-0103}
\and{C. Fremling}\inst{11,14}\orcidlink{0000-0002-4223-103X}
\and{A. Gangopadhyay}\inst{1}\orcidlink{0000-0002-3884-5637}
\and{K. Maguire}\inst{15}\orcidlink{0000-0002-9770-3508}
\and{G. Mo}\inst{12}\orcidlink{0000-0001-6331-112X}
\and{Y. Sharma}\inst{11}\orcidlink{0000-0003-4531-1745}
\and{N. Sravan}\inst{16}
\and{J. H. Terwel}\inst{15}\orcidlink{0000-0001-9834-3439}
\and{Y. Yang}\inst{9,17}\orcidlink{0000-0002-6535-8500}
}

\institute{The Oskar Klein Centre, Department of Astronomy, Stockholm University, AlbaNova, SE-10691, Stockholm, Sweden
\and{Graduate Institute for Advanced Studies, SOKENDAI, 2-21-1 Osawa, Mitaka, Tokyo 181-8588, Japan}
\and{National Astronomical Observatory of Japan, National Institutes of Natural Sciences, 2-21-1 Osawa, Mitaka, Tokyo 181-8588, Japan}
\and{The Oskar Klein Centre, Department of Physics, Stockholm University, AlbaNova, SE-10691, Stockholm, Sweden}
\and{Hamburger Sternwarte, University of Hamburg, Gojenbergsweg 112, D-21029 Hamburg, Germany}
\and{Center for Interdisciplinary Exploration and Research in Astrophysics (CIERA), Northwestern University, 1800 Sherman Ave.,
Evanston, IL 60201, USA}
\and{School of Physics and Astronomy, Monash University, Clayton, VIC 3800, Australia}
\and{University of North Carolina at Chapel Hill, 120 E. Cameron Ave., Chapel Hill, NC 27514, USA}
\and{Department of Astronomy, University of California, Berkeley, CA 94720-3411, USA}
\and{IPAC, California Institute of Technology, 1200 E. California Blvd, Pasadena, CA 91125, USA}
\and{Division of Physics, Mathematics and Astronomy, California Institute of Technology, Pasadena, CA 91125, USA}
\and{MIT Kavli Institute for Astrophysics and Space Research, 70 Vassar St., Cambridge, MA 02139, USA}
\and{Department of Physics, Lancaster University, Lancaster LA1 4YB, UK\    }
\and{Caltech Optical Observatories, California Institute of Technology, Pasadena, CA 91125, USA}
\and{School of Physics, Trinity College Dublin, The University of Dublin, Dublin 2, Ireland}
\and{Department of Physics, Drexel University, Philadelphia, PA 19104, USA}
\and{Department of Physics, Tsinghua University, Qinghua Yuan, Beijing 100084, China}
}

\abstract{Identifying the progenitors of thermonuclear supernovae (Type Ia supernovae; SNe Ia) remains a key objective in contemporary astronomy. The rare subclass of SNe Ia-CSM that interacts with circumstellar material (CSM) allows for studies of the progenitor’s environment before explosion, and generally favours single-degenerate progenitor channels. The case of SN Ia-CSM PTF11kx clearly connected thermonuclear explosions with hydrogen-rich CSM-interacting events, and the more recent SN 2020eyj connected SNe Ia with helium-rich companion progenitors. Both of these objects displayed delayed CSM interaction which  established their thermonuclear nature. Here we present a study of \sn, a Type Ia-CSM with delayed interaction. We analyse photometric and spectroscopic data that monitor the evolution of \sn and compare its properties with those of peculiar SNe Ia and core-collapse SNe. 
At early times, the evolution of \sn resembles a slightly overluminous SN~Ia. Later, the interaction-dominated spectra develop the same pseudocontinuum seen in Type Ia-CSM PTF11kx and SN 2020eyj. However, the later-time spectra of \sn lack hydrogen and helium narrow lines. Instead, a few narrow lines could be attributed to carbon and oxygen. We fit the pseudobolometric light curve with a CSM-interaction model, yielding a CSM mass of $1-2~$\msun. 
We propose that \sn was a Type Ia supernova that eventually interacted with a dense medium which was deficient in both hydrogen and helium. 
Whereas previous SNe Ia-CSM constitute our best evidence for nondegenerate companion progenitors, the CSM around \sn is more difficult to understand. We include a hydrodynamical simulation for a double-degenerate dynamical collision to showcase that such a progenitor scenario could produce significant amounts of hydrogen-poor CSM, although likely not as much as the inferred CSM mass around \sn. It is clear that \sn  challenges current models for stellar evolution leading up to a SN Ia explosion. 
}

\keywords{supernovae: general -- supernovae: individual: \sn, SN~2020eyj}

\maketitle

\section{Introduction} \label{sec:intro}
Supernovae (SNe) interacting with circumstellar material (CSM) are classified based on their spectra, which are dominated by narrow ($\lesssim10^{3}$~km~s$^{-1}$) emission lines \citep[e.g.][]{galyam2017}. These emission lines are interpreted as signs of shock interaction between fast-moving SN ejecta and slow-moving CSM \citep[see, e.g.][for a review]{smith2017}. The most commonly observed subtypes of interacting SNe are Type IIn \citep{Schlegel1990,nyholm2020} and Type Ibn \citep{ hosseinzadeh2017}, which have spectra dominated by hydrogen (H) and helium (He) emission lines, respectively. Recently, the classes of H/He-poor CSM-interacting stripped-envelope core-collapse supernovae (SNe Icn; \citealt{Galyam2021,perley2022}) and even a proposed SN Ien \citep{Schulze2025}, were also discovered.

A small fraction of thermonuclear SNe Ia also exhibits evidence of strong CSM interaction \citep{Silverman2013,Sharma2023}. These Type Ia-CSM SNe can show diluted SN Ia absorption features at peak brightness, but at later phases their spectra resemble those of SNe IIn with strong narrow H emission. SNe Ia-CSM therefore constitute some of the best candidates for the single-degenerate (SD) channel for SNe Ia, where a white dwarf (WD) accretes matter from a nondegenerate red giant, an AGB star, a He star, or a main-sequence companion star. In the alternative double-degenerate (DD) channel, where the SN Ia explosion would arise during the merger of two WDs, the CSM density should be much lower since the WDs must have lost their envelopes long ago \citep[e.g.][]{shen2013}.
 
A difficulty in the study of interacting SNe and their progenitors is that the CSM interaction typically dominates the spectral and light-curve evolution of the SN, and therefore conceals the underlying SN ejecta \citep{Leloudas2015}. Especially the subclass of SNe Ia-CSM and their spectroscopic relatives, SNe IIn, spurred a debate in the literature about the thermonuclear versus core-collapse nature of the former \citep[e.g.][]{hamuy2003,Benetti2006}. That debate surrounding SNe Ia-CSM was largely resolved with the discovery of PTF11kx \citep{dilday2012}. Whereas most SNe Ia-CSM have CSM interaction from the beginning, preventing an unambiguous classification, in PTF11kx the CSM interaction was delayed by $\sim 60$ days. Early-time spectra therefore allowed for a secure classification of PTF11kx as a SN Ia similar to SN 1991T \citep{Filippenko1992} and SN 1999aa \citep{Garavini2004}, which are slightly overluminous SNe Ia with weak \ion{Si}{II} absorption lines near maximum light. After the CSM interaction set in, PTF11kx revealed a strong, intermediate-width (1000 km~s$^{-1}$) H$\alpha$ emission line \citep{Silverman2013b,Graham2017}.

The discovery of Type Ia-CSM SN 2020eyj, which lacked narrow H lines in the late-time spectra, further argued for a population of thermonuclear progenitors with H-deficient companions \citep{Kool2022}. The unique optical observables of this object (no H but clear He lines) were also complemented with the first radio detection of a SN Ia, further strengthening the evidence for CSM interaction powering the late-time light curve. The discussions in \cite{Kool2022} argued for a progenitor system with a He-donor star, potentially similar to the V445 Puppis system \citep{Woudt2009}. 

In this paper, we report on an extensive follow-up campaign of \sn, which based on its peak spectrum was classified as a SN Ia, but where the later spectra and light-curve evolution clearly reveal substantial CSM interaction. \sn thus resembles both PTF11kx and SN 2020eyj in spectral and photometric evolution. The main difference is that the CSM around \sn is both He-poor and H-poor, in contrast to any previous SN Ia-CSM observed. 

This paper is organised as follows. In Section~\ref{sec:data} we describe the observations of \sn. Section~\ref{sec:analysis} includes analysis and discussion of the light-curve and spectral properties. In Section~\ref{sec:discussion} we fit the late light curve with a CSM interaction model and discuss the implications of our analysis. Section~\ref{sec:summary} summarises the findings. Throughout this paper we adopt a flat 
$\Lambda$CDM
cosmology with H$_0$ = 70 km~s$^{-1}$ Mpc$^{-1}$ and $\Omega_M = 0.3$.

\section{Observations}  \label{sec:data}
\subsection{Discovery, initial observations and classification}

\begin{figure}
\centering
    \includegraphics[width=9.0 cm]{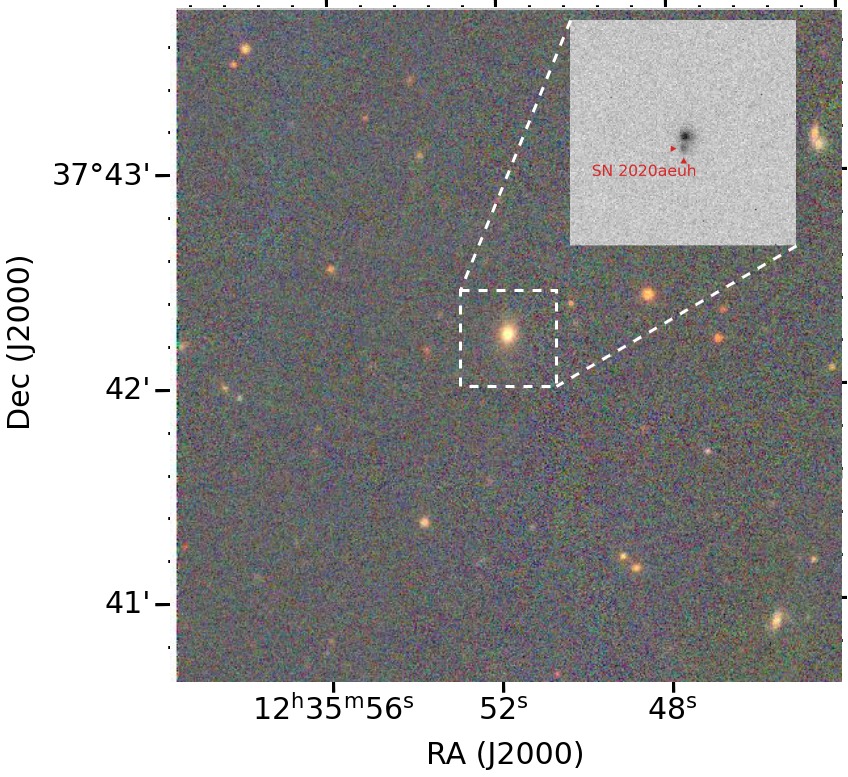}
    \caption{The field of \sn and its host galaxy. The colour image of the field is constructed from Pan-STARRS images in the \textit{gri} bands \citep{chambers2016}. The inset shows a negative image of the host galaxy, taken with P60 in the \textit{g}-band filter at an epoch of 197.8 days. The position of \sn is indicated with red arrows.
    }
    \label{fig:image}
\end{figure}

\sn (also known as ATLAS20bkbb, ZTF20acyroke) was first reported to the Transient Name Server (TNS\footnote{\href{https://www.wis-tns.org/}{https://www.wis-tns.org/}}) in January 2021 by the Asteroid Terrestrial-impact Last Alert System (ATLAS) collaboration \citep{discoveryatlas}, with their first detection from Christmas day 2020 December 25 (UTC dates are used throughout this paper; $t_{\mathrm{first}}^{\mathrm{ATLAS}}$ = 59208.6~MJD) at $19.27 \pm 0.13$ mag in the cyan band.

\sn was also observed with the Palomar Schmidt 48-inch (P48) Samuel Oschin telescope as part of the Zwicky Transient Facility (ZTF) survey \citep{Bellm2019,graham2019a}. Forced photometry revealed a first detection ($>3\sigma$) in the $r$ band already on 2020 December 16 ($t^{\textrm{ZTF}}_{\textrm{first}}$  = 59199.415 MJD), at the J2000.0 coordinates 
RA $=12^{h}35^{m}51.86^{s}$, Dec. $=+37\degr42\arcmin12.4\arcsec$, with a host-subtracted brightness of $20.83 \pm 0.34$~mag. ZTF yielded a nondetection at the $3\sigma$  level on the preceding night, with ZTF ($5\sigma$) limiting magnitudes in the $g$ and $r$ bands of 20.9 and 20.6~mag, respectively. Phases (in rest-frame days) in this paper will be relative to the first detection by ZTF ($t_{0}$~=~$t^{\mathrm{ZTF}}_{\mathrm{first}}$), which is also close to the explosion date, given the constraints from the preceding nights.

\sn (Fig.~\ref{fig:image}) was classified as a Type Ia SN at redshift $z=0.126$ \citep{weil2021} based on a spectrum obtained on 2021 January 13 with the Ohio State Multi-Object Spectrograph (OSMOS) on the 2.4\,m Hiltner telescope, publicly available on the TNS. This is the only spectrum of the early SN Ia phase. A second spectrum was not obtained until 64 days later, after the light curve had started brightening again. This second spectrum was obtained with the Alhambra Faint Object Spectrograph and Camera (ALFOSC) on the Nordic Optical Telescope \citep[NOT;][]{djupvik2010}, and showed a transformed SN, with a spectrum dominated by CSM interaction features (see details in Sect.~\ref{sec:latespectra}). 

We obtained a redshift of $z = 0.1265 \pm 0.0028$ by measuring host-galaxy absorption lines in a later NOT spectrum, which we will use throughout this paper. This is well into the Hubble flow, thus we will use the redshift-derived luminosity distance of 595 Mpc ($\mu = 38.944$~mag).

\subsection{Photometry}

Following the discovery, we obtained regular follow-up photometry with P48 until \sn faded below the detection limit, 180 days after the first detection. Initial follow-up observations were performed in the $g$ and $r$ bands, with later phases also covered in the $i$ band. This was accomplished with the ZTF camera \citep{dekany2020} on the P48. The $g$-, $r$-, and $i$-band P48 data include points on 88, 86, and 23 separate nights, respectively.

The photometry from the P48 come from the ZTF forced-photometry service \citep{masci2019, Masci2023} and from the P60 using the pipeline described in \cite{fremling2016}. The photometric magnitudes of \sn are listed in Table~\ref{tab:photometry}. Additionally, we make use of the forced-photometry service from the ATLAS survey \citep{Tonry2018,2020PASP..132h5002S, ATLAS_FP}, which contains valuable photometry in the $o$ and $c$ bands.

\begin{figure*}
\centering
\sidecaption
    \includegraphics[width=12cm]{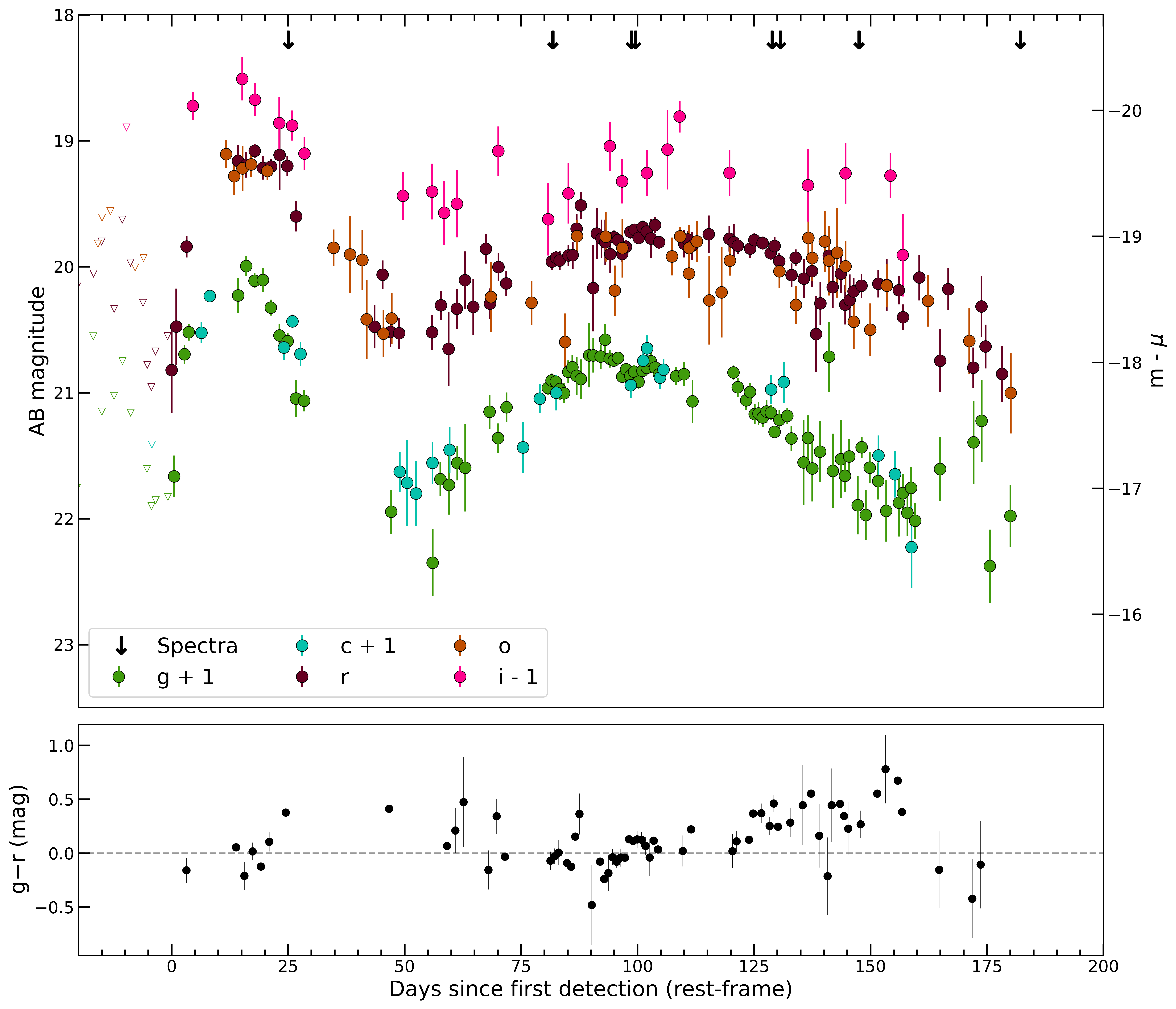}
    \caption{Light curves of \sn, corrected for MW extinction, with phases in rest-frame days since first detection. Early nondetections with 5$\sigma$ upper limits are indicated by downward-facing empty triangles. Apparent magnitudes are provided on the left-hand axis and absolute magnitudes to the right. The photometry has been binned into one-night bins. The arrows on top indicate the epochs of spectroscopy. The lower panel shows the $g - r$ colour evolution.}
    \label{fig:lightcurve}
\end{figure*}

The extinction corrections are applied using the \citet{cardelli1989} extinction law with $R_V=3.1$. The extinction in the Milky Way (MW) was obtained from \cite{Schlafly2011} as $E(B-V) = 0.0156$ mag. We do not correct for host-galaxy extinction, since there is no sign of \ion{Na}{i d} absorption lines in our spectra. The ATLAS and P48 light curves are shown in Fig.~\ref{fig:lightcurve}, binned into 1-night bins to enhance the signal-to-noise ratio (S/N).

\subsection{Spectroscopy}

The public SN Ia classification spectrum of \sn mentioned above was obtained 24 days after discovery \citep{weil2021}. The SN never quite reached apparent magnitude 19, which is the cutoff for the Bright Transient Survey \cite[BTS;][]{Fremling2020,perley2020b}. It was thus
too faint for spectral monitoring with the P60 Spectral Energy Distribution Machine \citep[SEDM;][]{blagorodnova2018, rigault2019} which is typically the workhorse for the BTS project.
Instead, once the SN had rebrightened, we obtained four epochs of spectroscopy with the NOT using ALFOSC and three epochs with the Keck I telescope and the Low-Resolution Imaging Spectrometer \citep[LRIS;][]{Oke+1995}. A log of the spectral observations is provided in Table~\ref{tab:spec}. The epochs of spectroscopy are also indicated by the downward-pointing arrows on top of the light curves in Fig.~\ref{fig:lightcurve}. The spectra obtained with ALFOSC and LRIS were reduced with the \texttt{Pypeit} \citep{Prochaska2020} and \texttt{LPipe} \citep{perley2019} software packages, respectively. The spectra were absolute calibrated against the $r$-band magnitudes using Gaussian Process (GP) interpolated magnitudes and then corrected for MW extinction. All spectral data are made available via WISeREP\footnote{\href{https://www.wiserep.org/}{https://www.wiserep.org/}} \citep{wiserep}. We present the peak classification spectrum in Fig.~\ref{fig:classificationspectrum} and the later sequence of spectra in Fig.~\ref{fig:laterspectrasequence}.

\section{Analysis} \label{sec:analysis}

\sn is a rare example of a SN Ia showing delayed CSM interaction. Only three delayed Type Ia-CSM SNe with extended optical light curves have previously been reported in the literature: SN~2002ic \citep{hamuy2003,woodvasey2004}, PTF11kx \citep{dilday2012}, and Type Ia-CSM-He SN~2020eyj \citep{Kool2022}, but see also the discussion in \citet[][their Sect. 4.1]{Sharma2023}. None of these SNe showed light curves with the SN Ia and CSM phases as clearly separated as we see for \sn, nor did they show a clear optical rebrightening \citep[a handful of SNe Ia have shown late-time rebrightening in the mid-infrared due to CSM interaction;][]{Mo2024}. The first peak phase of \sn is consistent with the diffusion peak of a radioactively powered Type I SN. The second phase of the light curve starts at 
$\sim 50$ days, when the SN increases in brightness again over the course of another $\sim 50$ days, followed by a slow decline. 

In this section we will first explore the early SN Ia classification spectrum, by comparing it to template spectra of SNe Ia and SNe Ibc included in \texttt{SuperNova} \texttt{IDentification}\footnote{\href{https://people.lam.fr/blondin.stephane/software/snid/}{https://people.lam.fr/blondin.stephane/software/snid/}} \citep[\texttt{SNID};][]{blondin2007}. Next,  we investigate the properties of the late-time spectra and light curves of \sn in the ``CSM phase,'' and compare \sn to various CSM-interacting SNe from the literature, including PTF11kx and SN~2020eyj.

\subsection{Classification spectrum}
\begin{figure}
\centering
    \includegraphics[width=\hsize]{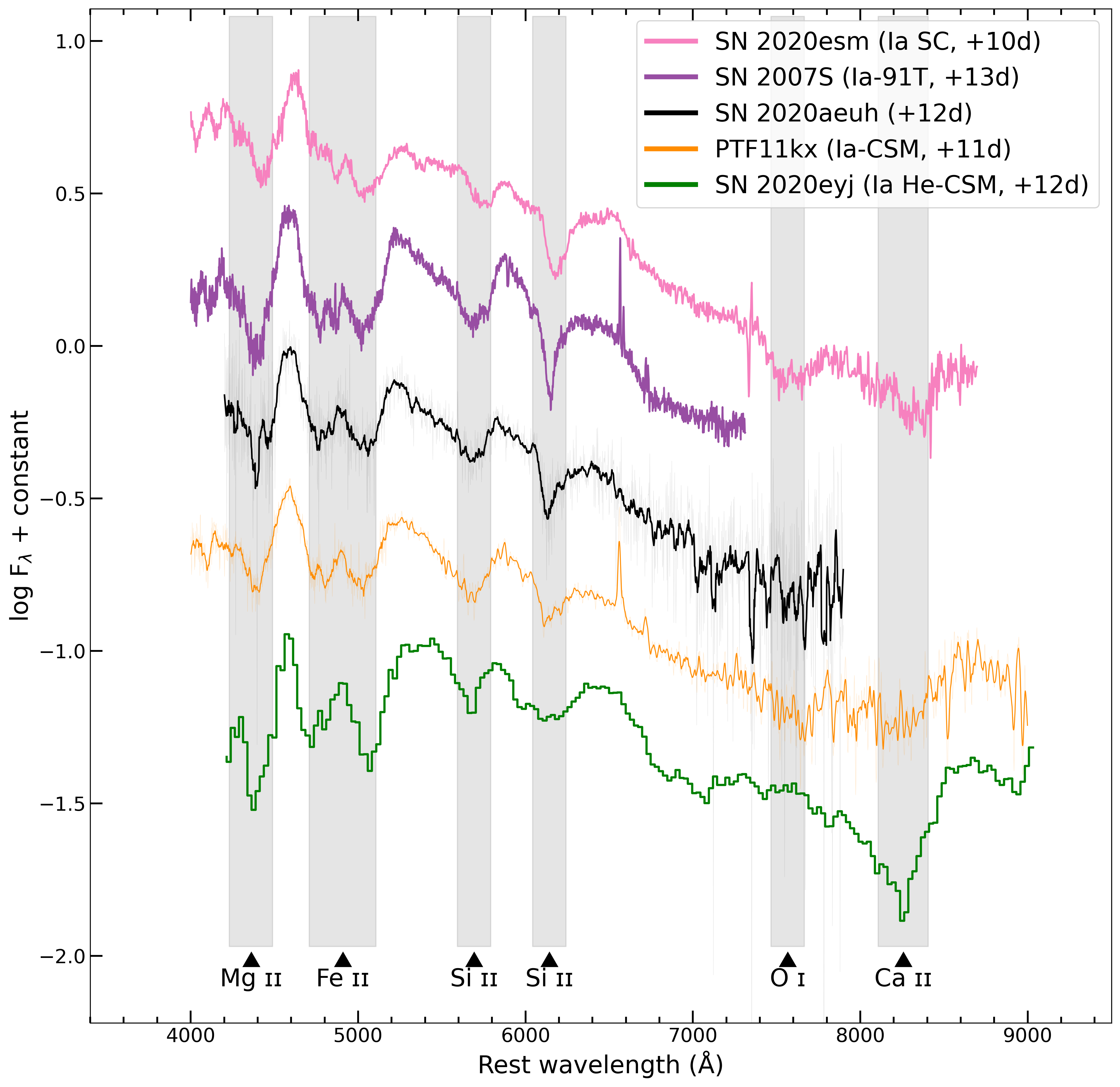}
    \caption{Comparison of the classification spectrum of \sn (in black) with a variety of SNe Ia from the literature (see text for discussion and references). The locations of some common transitions are provided at the bottom.}
    \label{fig:classificationspectrum}
\end{figure}

The first, and only, spectrum obtained during the peak phase of \sn is characterised by broad absorption features (Fig.~\ref{fig:classificationspectrum}), based on which the SN was classified as a Type Ia \citep{weil2021}. Using \texttt{SNID}, supplemented with the Type Ibc SN templates from \cite{Modjaz2014}, we obtain good (rlap $> 12$) matches with SN Ia templates, in particular those of subtype Ia-91T. The best core-collapse SN template matches are of Type Ic, such as SN~1994I, with rlap $\approx 7$. The phases of the matched SN Ia templates are around 10--15 days post peak, whereas the matched SN Ic templates are all within a few days of peak brightness. The peak epoch of \sn in the $g$ band is $t_{peak}$ = 59213.4 MJD (Sect.~\ref{sec:lightcurve}), which puts the phase of the classification spectrum at +12 days post peak. The classification spectrum's phase is consistent with the phases of the SN Ia templates, but inconsistent with any of the matched SN Ic templates.

The spectrum of \sn is well matched in particular to those of Type Ia-91T SNe, for example SN~2007S \citep{Blondin2012}. 
SNe Ia-91T have previously been connected to SNe Ia-CSM \citep{Leloudas2015,Phillips2024}. They
typically have weak \ion{S}{II} and \ion{Si}{II} features near maximum light, and are photometrically overluminous compared to normal SNe Ia. These characteristics are shared with SNe Ia-CSM 
such as SN Ia-CSM PTF11kx and SN~2020eyj,  and also with so-called super-Chandrasekhar (SC) SNe Ia such as
SN~2020esm \citep{Dimitriadis2022}. In Fig.~\ref{fig:classificationspectrum} we show a comparison of these SNe Ia with the classification spectrum of \sn at similar phases post peak. 
We thus confirm the classification of \sn as a SN Ia in concordance with the initial report by \cite{weil2021}.

\subsection{Late-time spectra}\label{sec:latespectra}

\begin{figure*}
\centering
    \includegraphics[width=\hsize]{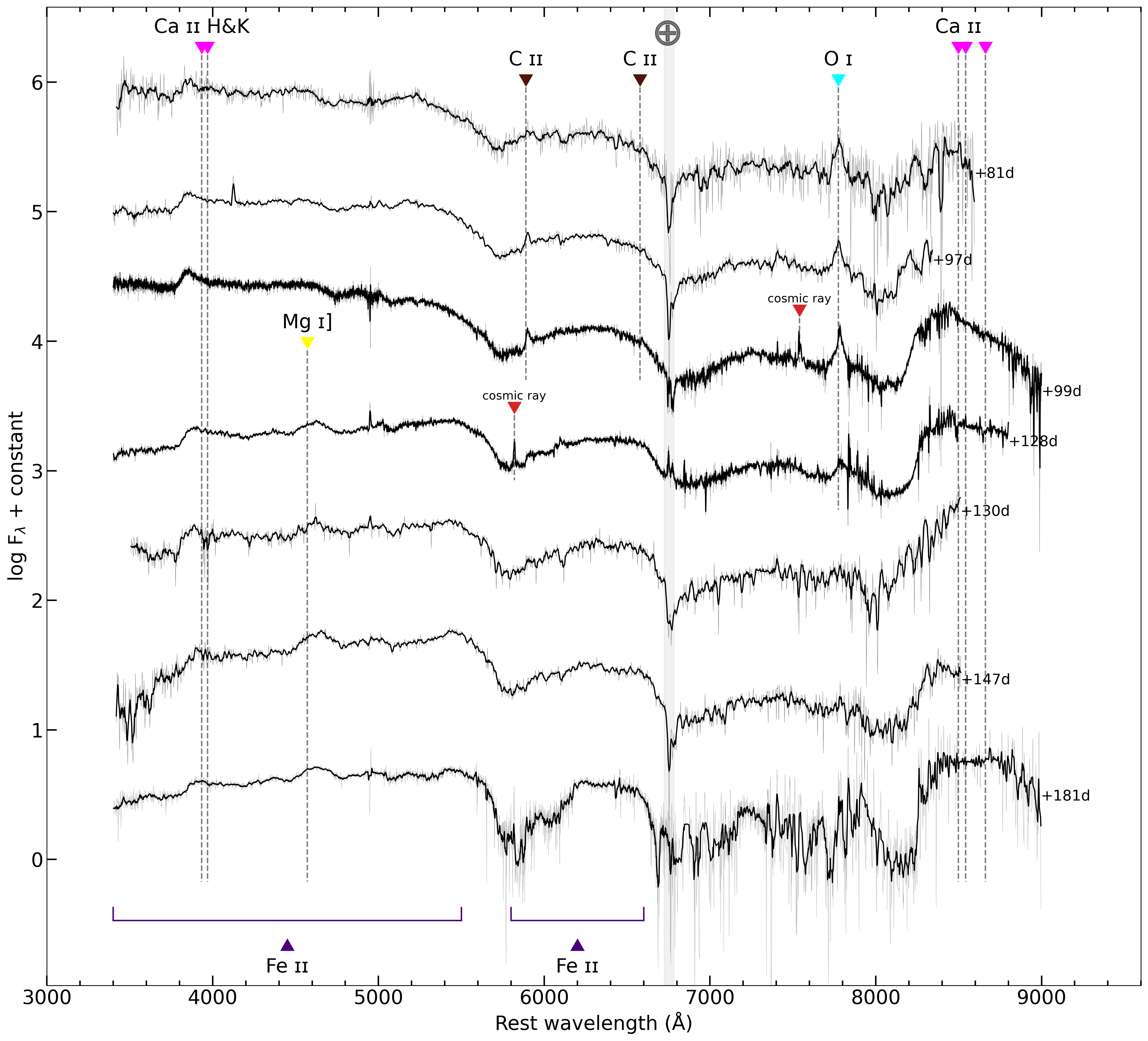}
    \caption{Spectral sequence of \sn at late phases displaying our 7 spectra between 81 and 181 days past first detection, when the SN is clearly dominated by CSM interaction. The smoothed spectra (using a Savitzky–Golay filter) are shown in solid lines while the original in semi-transparent. The spectral evolution is rather slow during these 100 days. The spectrum with best signal, at 99 days, is compared to spectra of other SNe in Fig.~\ref{fig:csm_spectrum_comparison}.}
    \label{fig:laterspectrasequence}
\end{figure*}

The series of late-time spectra is shown in Fig.~\ref{fig:laterspectrasequence}. The first of these was obtained using ALFOSC (Table~\ref{tab:spec}) at +81 days, only after it was realised that \sn was not a normal SN Ia. 

\begin{figure}
\centering
    \includegraphics[width=\hsize]{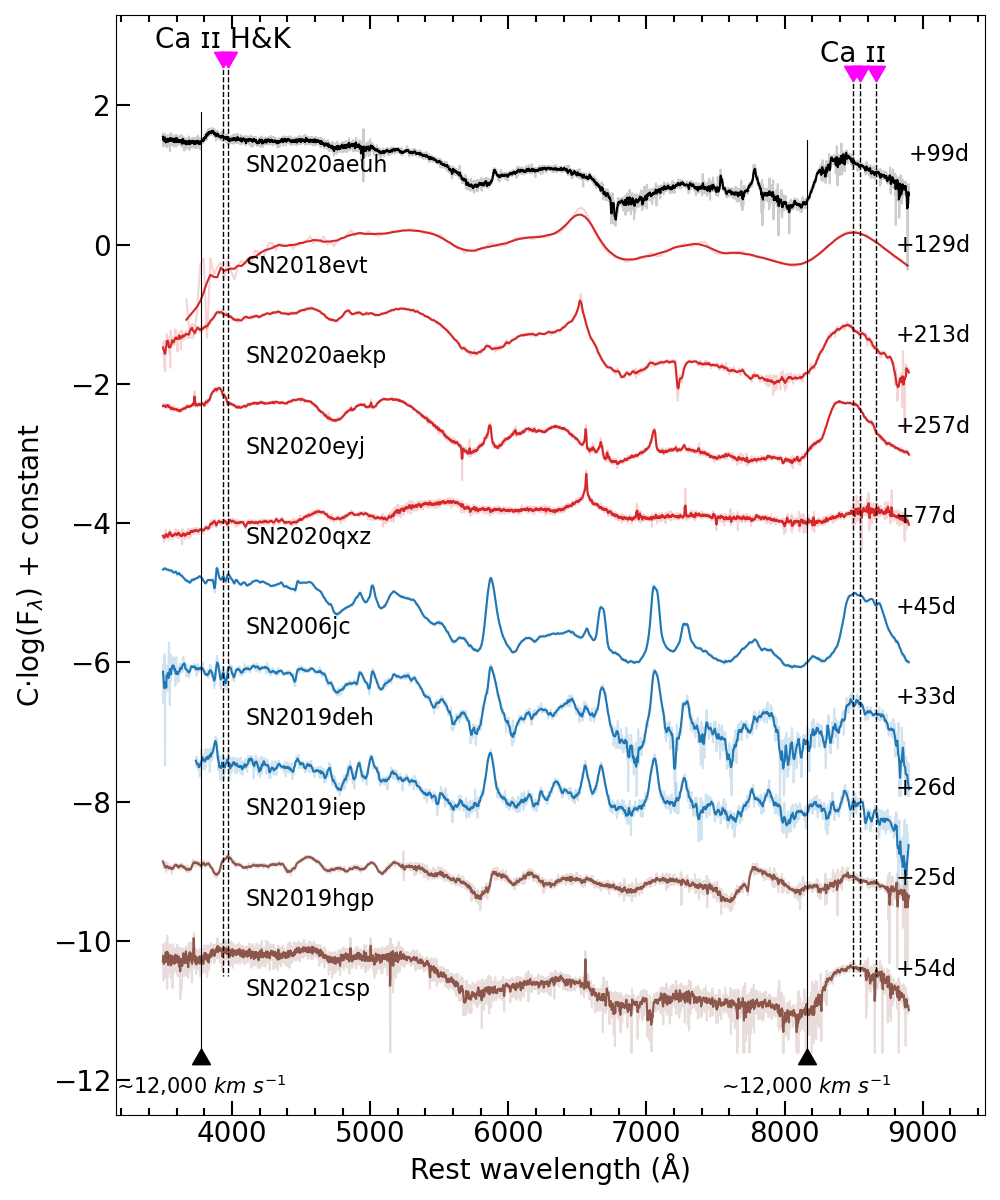}
    \caption{A comparison of the spectrum of \sn at +99 days with a selection of other interacting Type I SNe. The shown Type Ia-CSM (in red) are SNe 2018evt, 2020aekp, 2020qxz \citep{Sharma2023}, and SN 2020eyj \citep{Kool2022}. Type Ibn (blue): SN 2006jc \citep{foley2007}, SN 2019deh, and SN 2019iep. Type Ic (brown): SN 2019hgp \citep{Galyam2021} and SN 2021csp \citep{perley2022}. All spectra have been shifted and multiplied by a numerical factor for visual purposes.  Phases are relative to first detection. Dashed lines mark the calcium features. Solid lines indicate the calcium velocity offset estimated for \sn.}
    \label{fig:csm_spectrum_comparison}
\end{figure}
The 7 late-time spectra show little evolution, and are characterised by broad emission features typical of interacting SNe (Fig.~\ref{fig:csm_spectrum_comparison}), such as the steep pseudocontinuum blueward of $\sim 5700$~\AA, attributed to a forest of blended \ion{Fe}{ii} lines \citep{chugai2009,smith2012,Stritzinger2012,pastorello2015IV}. We also attribute the pseudocontinuum hump at $\sim 6000$ to $\sim 6800$~\AA\ to a blend of \ion{Fe}{ii} lines \citep[see, e.g. the spectral modelling of interacting SNe Ibn in][]{Dessart2022A&A...658A.130D}. The spectra of \sn also show a broad \ion{Ca}{ii} near-infrared triplet, which is commonly observed in SNe Ia-CSM \citep{Silverman2013}, as well as a feature that likely corresponds to \ion{Ca}{ii h\&k}.

Aside from the features mentioned above, the late-time spectra lack the narrow or intermediate-width emission lines that typically arise from interaction of SN ejecta with CSM. In particular, the highest S/N spectrum of \sn, obtained at +99 days, does not show any evidence of narrow  H or He emission. This is unprecedented in a SN Ia-CSM, and somewhat reminiscent of the late-time spectra of SNe Icn \citep{Galyam2021}. The spectrum shows only two clear narrow emission features: a prominent line at an observed wavelength of $8767 \pm 8$~\AA, and a narrow emission line at $6647 \pm 10$~\AA. At the host redshift of $z = 0.1265$ this translates to rest wavelengths of $7783 \pm 8$~\AA\ and $5900 \pm 9$~\AA, respectively.

We suggest that the longest wavelength emission feature is due to \ion{O}{i} $\lambda$7774, which would imply a velocity offset of the SN from the host redshift of $\sim 350$~km~s$^{-1}$.Assuming such a velocity offset, the other narrow emission feature would reside at $\sim 5893$~\AA, consistent with \ion{C}{ii} $\lambda$5890. That line identification is corroborated by a weak emission feature at $6584 \pm 6$~\AA\ (with $z = 0.1265$ and 350~km~s$^{-1}$), which might be interpreted as \ion{C}{ii} $\lambda$6578. The rare subclass of core-collapse SNe (CCSNe) that interact with a He/H-free CSM, Type Icn SNe, have spectra that exhibit narrow (P-Cygni) \ion{C}{ii} emission \citep{Davis2022}, albeit more typically at epochs within a few days post peak. We do note that \ion{C}{ii} $\lambda\lambda$7231, 7236 is not observed in any of the spectra of \sn, while this line can be of equal strength to \ion{C}{ii} $\lambda$5890 in SNe Icn. Alternatively, the 5900~\AA\ line could be due to \ion{Na}{i d} $\lambda\lambda$5890, 5896. We disfavour \ion{He}{i} $\lambda$5876, as this would require a velocity shift of 1200~km~s$^{-1}$, which is not consistent with the 
\ion{O}{i} $\lambda$7774 identification, and no other He lines such as He I $\lambda$7065 are observed.

We examined the three narrow features in the LRIS spectrum with the highest S/N, at 99 days. Both the $\sim 5893$~\AA\ and $\sim 6584$~\AA\  lines have symmetric profiles. This allows for a good Gaussian fit, yielding a full width at half-maximum intensity (FWHM) of $1263 \pm 108$~km~s$^{-1}$ 
and $748 \pm 220$~km~s$^{-1}$ 
for the suggested \ion{C}{ii} $\lambda \lambda$5890, 6578, respectively. The most pronounced line at $\sim$7783~\AA\ shows a blue-wing excess. A Lorentzian fit gives a FWHM of $1343 \pm 153$~km~s$^{-1}$.

The spectral sequence of \sn shows little to no evolution. The major change involves the disappearance of the narrow features. This is first observed around the +128~d spectrum and onward (Fig.~ \ref{fig:laterspectrasequence}). Simultaneously, at 128 days we observe the emergence of a broad feature within the blue quasicontinuum. This broad line is centred at $4624 \pm 5$~\AA\ in the rest frame. A potential identification would be \ion{Mg}{i]} $\lambda$4571. In that case, the offset of $\sim 350$~km~s$^{-1}$ from the host's redshift is not enough to explain the redshift of the feature. The appearance of a red shoulder of \ion{Mg}{i]} could be a visual effect owing to the blending of the line with the numerous iron lines in the blue quasicontinuum.

The main difference between the late-time spectra of \sn and those of other Type Ia-CSM SNe is thus the nature of the emission lines. Normal SNe Ia-CSM show strong H$\alpha$ and no or only weak \ion{He}{i} emission \citep{Sharma2023}. SN~2020eyj did not exhibit Balmer emission, but rather multiple strong \ion{He}{i} lines. For \sn, both H and He lines are missing.

\subsection{Light curve}\label{sec:lightcurve}

The light curve of \sn consists of two distinct phases (Fig.~\ref{fig:lightcurve}). During the first phase, the light curve follows a typical bell-shape, which is observed to peak in both the $g$ and $r$ bands around $M = -19.7$ at 15 days. After the peak, the light curve declines and reddens until $\sim40$--50 days, after which a conspicuous rebrightening sets in. 

This rebrightening is mostly pronounced at shorter wavelengths, resulting in a colour change to the blue (at $\sim50$--100~d) before 
 reddening again (at $\sim 100$--170~d) during the second decline (Fig.~\ref{fig:lightcurve}, lower panel). A similar colour evolution was observed in the interaction-dominated phase of the Type Ia-CSM-He SN~2020eyj. The development of the blue pseudocontinuum is responsible for the colour evolution during the second peak of \sn, as it overlaps with the ZTF $g$-band filter \citep{Kool2022}. As shown in the spectral sequence of \sn (Fig.~\ref{fig:laterspectrasequence}), the blue pseudocontinuum is prominent at +99~d, and then gradually decreases in strength over time, accounting for the rereddening after $\sim 100$~d. 

\begin{figure}
\begin{center}
	    \includegraphics[scale=0.25]{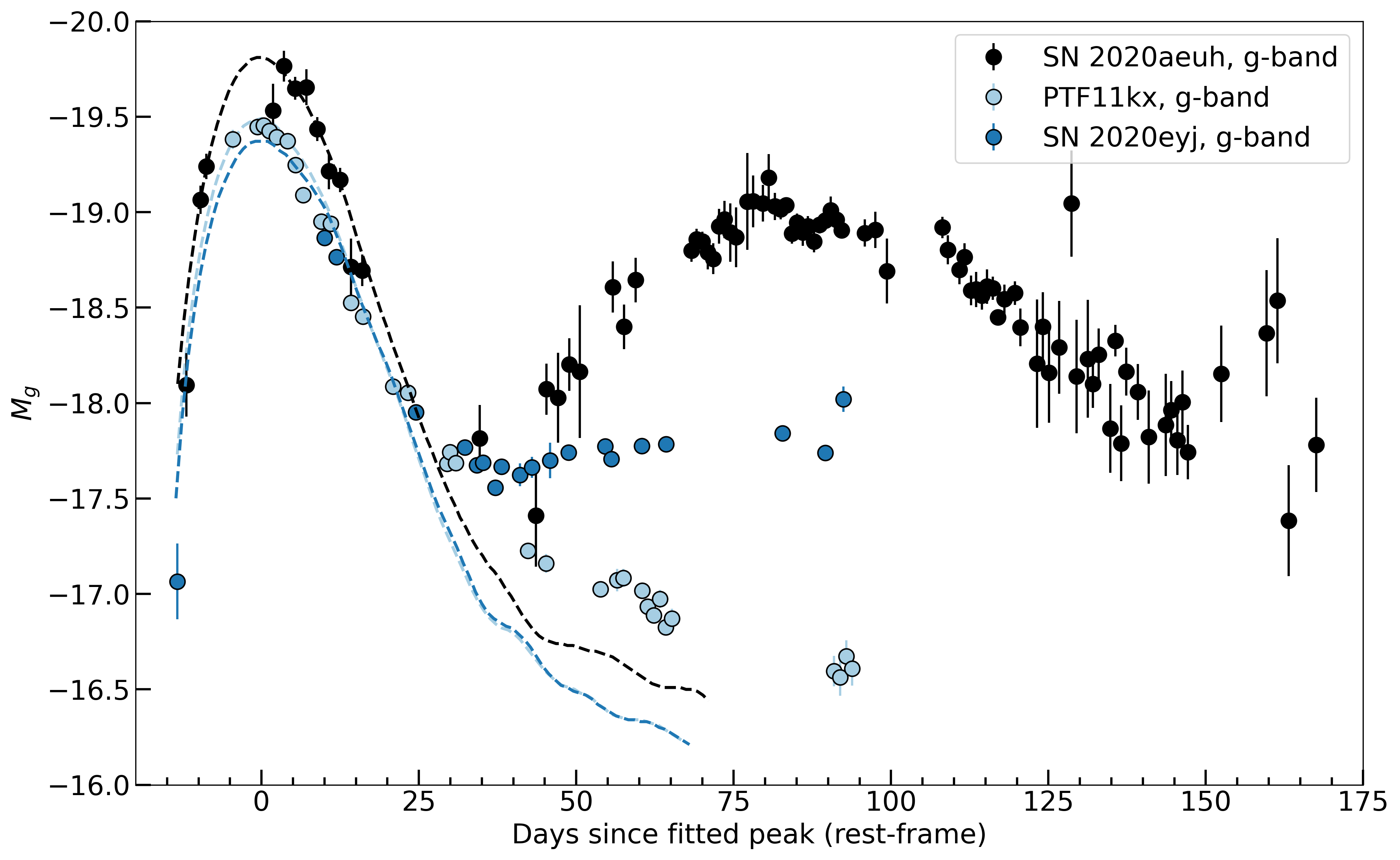} 
		\includegraphics[scale=0.25]{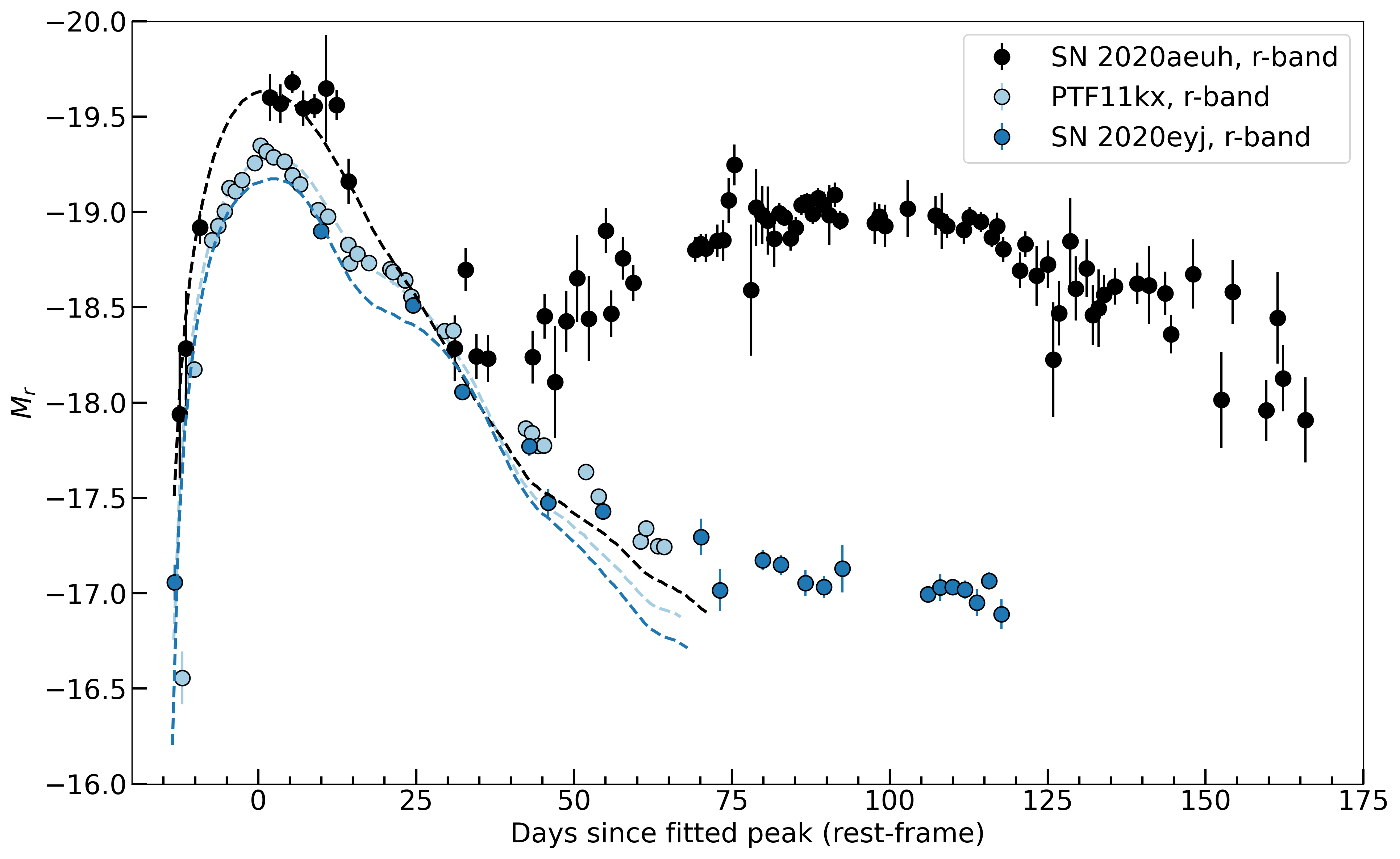}
\end{center}
	\caption{The $g$- (upper panel) and $r$-band (lower panel) light curves of the delayed SNe Ia-CSM PTF11kx (light blue), SN~2020eyj (dark blue), and \sn (black). The comparison light curves are corrected for the host-galaxy extinction inferred with \texttt{SNooPy}. The modelled light curves are overplotted. All three SNe evolve in a similar way up until $\sim25$ days after peak brightness, after which the light curves diverge, in particular in the  $g$ band. \sn shows a much more conspicuous late-time CSM interacting light curve that even increases in luminosity.}
	\label{fig:lightcurve_comparison}
\end{figure}

Based on the spectra, we have established that \sn is consistent with a SN Ia during the first peak, and is dominated by CSM interaction during the second peak (Sect.~\ref{sec:latespectra}). We fit the $g$, $r$, and $i$ photometry with SN Ia template light curves using \texttt{SNooPy}\footnote{\href{https://users.obs.carnegiescience.edu/cburns/SNooPyDocs/html/}{https://users.obs.carnegiescience.edu/cburns/SNooPyDocs/html/}} \citep{Burns2011} to determine the peak of the light curves in the different filters, and to see if the stretch of the light curve is consistent with the peak luminosity in accordance with the Phillips relation \citep{phillips1993} for SNe Ia. The light curves (Fig.~\ref{fig:lightcurve_comparison}) are well fit by a model with a stretch of $1.19 \pm 0.04$ corresponding to an expected peak magnitude 0.15 mag fainter than what is observed, which is well within the range for normal Type Ia SNe. 
We also note that \sn was investigated in connection with the second data release of cosmological SNe Ia from ZTF \citep{Rigault2025,Dimitriadis2025, Terwel2025}. It yielded a good SALT2 fit with parameters\footnote{\href{http://ztfcosmo.in2p3.fr/target/ZTF20acyroke}{http://ztfcosmo.in2p3.fr/target/ZTF20acyroke}} $M_{g} = -19.85 \pm 0.08~$mag and $\Delta m_{15,g} = 0.79 \pm 0.15~$mag ($x_1 = 0.92 \pm 0.26$, $c = -0.05 \pm 0.04$).
SNe Ia-CSM have peak magnitudes that are typically brighter than those of normal SNe Ia, in the range $−21.3  \leq M_R \leq -19$~mag \citep{Silverman2013,Sharma2023}. The $r$-band light-curve model of \sn peaks at $M_r = -19.7 \pm 0.1$~mag, consistent with the subtype.

From the light-curve fits we infer a peak epoch of $t_{\rm peak}$ = 59213.4 $\pm$ 0.3 MJD, in the $g$ band. If we assume the explosion epoch of \sn to be halfway between the last nondetection and the first detection, at 59198.9 $\pm$ 0.5 MJD, this results in a rise-time lower limit of $12.5 \pm 0.5$ days. This is fast for a SN Ia; normal SNe Ia typically have longer rise times \citep[$18.5 \pm 1.6$ d;][]{Miller2020}. Interestingly, both PTF11kx and SN~2020eyj also had unusually fast rise times, at $\sim 14$ days in the $g$ band \citep{Kool2022}. 

\subsection{Pseudobolometric light curve}\label{sec:pbolometric}

To construct a bolometric light curve, we first interpolated the light curves in each band (\textit{gcroi}) to fill in the missing epochs. The $g$ band has the most populated light curve and was used as the reference band. A polynomial interpolation of order 6 was chosen to estimate the magnitudes of the individual-band light curves at the missing epochs. The polynomial's order is high enough to capture the two peaks in the light curve and low enough to not significantly overfit the data. 
We approximated a K-correction by transforming the filters' effective wavelengths and zero-point fluxes to the rest frame \citep[following the procedure in \texttt{SUPERBOL};][]{Nicholl2018} and converted magnitudes to flux densities ($F_{\lambda}$). Once we obtained a spectral energy distribution (SED) consisting of observed and interpolated data points at each epoch, we were able to construct a pseudobolometric light curve employing different assumptions.

First, we integrated the SED across the observed bands (\textit{gcroi}), enforcing the flux density to go to zero at the edges of the outer filters, using the \texttt{SuperBol} package \citep{Nicholl2018}. This bolometric light curve peaks at $\sim1.04\times10^{43}$~erg~s$^{-1}$. Obviously, this pseudobolometric light curve can only serve as a lower limit on the true bolometric light since it ignores the contribution from the near-ultraviolet (NUV) and near-infrared (NIR). \cite{Lyman2014} showed that in the case of stripped-envelope CCSNe (SE-CCSNe), the flux contribution of the optical regime alone (from $U$ to $K$ band) is $\sim 50$\%. We can expect the contribution of our optical SED to be even less given that we have narrower wavelength coverage.

Next, we fitted a blackbody spectrum to the observed SEDs and used the best fit to extrapolate the flux densities in the NUV (2000~\AA\ -- $\lambda_{\mathrm{eff,g}}$) and NIR ($\lambda_{\mathrm{eff,i}}$ -- 60,000~\AA), using \texttt{SuperBol}. 
The bolometric light curve constructed in this way peaks at $\sim8.75\times10^{43}$~erg~s$^{-1}$. We consider this to be an upper limit for the true bolometric peak luminosity, since for the NUV contribution, a blackbody assumption often overestimates the flux. This is because we can expect significant line blanketing during the early part of the light curve. At later epochs, for the interaction-dominated light curve, we observe the blue pseudocontinuum driving the colour evolution of \sn (Sect.~\ref{sec:lightcurve}). This also causes the NUV part to deviate from the blackbody approximation.

In order to relax the blackbody assumption but still include some contribution from the NUV, we linearly extrapolated the flux in that regime. Following the methodology of \cite{Lyman2014}, we assumed zero flux at 2000~\AA\ and linearly extrapolate the flux between 2000~\AA\ and $\lambda_{\mathrm{eff, u}}$. Our observations extend until the $g$ band, so we assumed the flux in the $u$ band to be equal to the flux in the $g$ band at every epoch. The NIR was again extrapolated as a blackbody tail until 60,000~\AA. This bolometric light curve peaks at $\sim2.45\times10^{43}$~erg~s$^{-1}$, a factor of 2.4 higher than the lower limit. We use this pseudobolometric light curve in the subsequent analysis.

Finally, we compared our adopted pseudobolometric light curve with one constructed using the \citet[][their Eq. 12 and Table 2]{Lyman2014} pseudobolometric corrections for SE-CCSNe. The bolometric light curve constructed for \sn using the Lyman-corrections for the $g-r$ colour differs from our pseudobolometric light curve by only $\sim 0.04$~dex. 

\subsection{$^{56}$Ni mass estimation}
We can estimate the amount of $^{56}$Ni produced in \sn from its pseudobolometric light curve using the Arnett rule \citep{arnett1982},
which states that the peak bolometric luminosity equals the instantaneous total energy production rate from the radioactive decay of $^{56}$Ni, $L_{\rm peak} = \epsilon(t_{\rm peak})$. \cite{Nadyozhin1994} calculated $\epsilon(t)$ to be 
\begin{equation}
    \epsilon(t) = [6.45\times 10^{43} e^{-t/\tau_{\rm Ni}}+ 1.45\times 10^{43}e^{-t/\tau_{\rm Co}}]\frac{M_{\rm Ni}}{\rm M_{\odot}}~\rm erg~ \rm s^{-1}\, ,
    \label{eq:epsilon}
\end{equation}
where $\tau_{\rm Ni} = 8.8~$d and $\tau_{\rm Ni} = 111.3~$d are the $e$-folding timescales for the $^{56}$Ni and $^{56}$Co decays. Using $L_{\rm peak}\approx 2.45\times10^{43}~\rm erg~\rm s^{-1}$ and $t_{\rm peak}\approx 12.5$~days, we obtain $M_{\rm Ni}\approx 0.86$~\msun. This is large for a normal SN Ia \citep[e.g.][]{Stritzinger2006A&A...460..793S,Bora2022}, but  is expected for a luminous SN 1991T-like object \citep{Nugent1995, Hoeflich1996}. There are also several caveats in this estimate, from the simple method of estimating the mass, the uncertainty of the bolometric light curve encapsulating all the flux, and any contribution from the circumstellar 
interaction to the first peak.

\subsection{Host Analysis}\label{sec:host}

We retrieved science-ready coadded images of the host galaxy of \sn from the Sloan Digital Sky Survey DR 9 (SDSS; \citealt{Ahn2012a}) and processed \textit{WISE} images \citep{Wright2010a} from the unWISE archive \citep{Lang2014a}\footnote{\href{http://unwise.me}{http://unwise.me}}. The unWISE images are based on the public \textit{WISE} data and include images from NEOWISE-Reactivation mission R-7 \citep{mainzer2014, Meisner2017a}. We used the software package \texttt{LAMBDAR} (Lambda Adaptive Multi-Band Deblending Algorithm in R) \citep{Wright2016} and tools presented by \citet{Schulze2021} to measure the brightness of the host galaxy. The host is well detected in all bands. We measure a brightness of $18.01 \pm 0.03$ mag in the SDSS $r$ band (corrected for MW reddening). The brightness in other filters is reported in Table \ref{tab:host}.

The SED was modelled with the software package \texttt{Prospector}\footnote{\href{https://github.com/bd-j/prospector}{https://github.com/bd-j/prospector}} version 0.3 \citep{Johnson2021a}. We assumed a linear-exponential star-formation history, the \cite{Chabrier2003a} initial mass function, the \cite{Calzetti2000a} attenuation model, and the \cite{Byler2017a} model for the ionised gas contribution. The priors were set as described by \cite{Schulze2021}. Based on these assumptions, we measure a k-corrected absolute magnitude of $M_r=-20.94 \pm 0.05$~mag (not corrected for host reddening) and living-star mass of $\log\,M_\star/\mathrm{M}_\odot=10.50^{+0.13}_{-0.18}$. The SED fit reveals no sign of recent star-formation activity. The formal $3\sigma$ limit is $<0.37~M_\odot\,\rm yr^{-1}$. The absence of H$\alpha$ emission in the Keck spectrum from +181 days allows us to tighten the limit further. We fit a Gaussian line profile at the expected location of H$\alpha$ and fixed its FWHM to the FWHM resolution of the LRIS 400/8500 grating for an $1\farcs0$-wide slit. This yields a flux limit $5.4\times10^{-18}~\rm erg\,cm^{-2}\,s^{-1}$ at the $3\sigma$ confidence level. Using the \cite{Kennicutt1998a} relation between the H$\alpha$ luminosity and star-formation rate (SFR) and the correction term from \cite{Madau2014a} to convert from a Salpeter to the Chabrier IMF, we infer an SFR limit of $<10^{-3}~\rm M_\odot\,\rm yr^{-1}$.

\begin{figure}
\centering
    \includegraphics[width=\hsize]{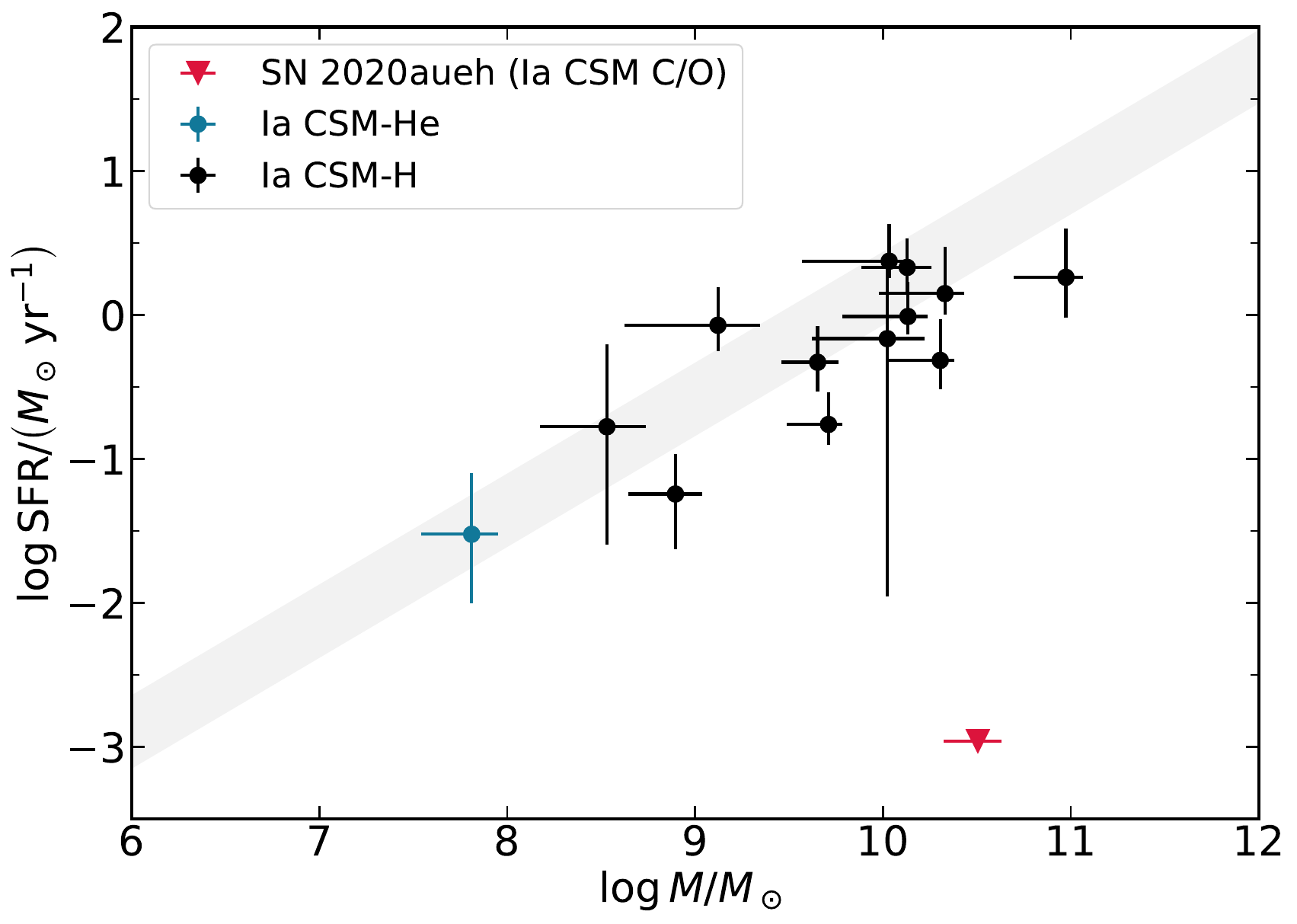}
    \caption{Star-formation rate and stellar mass of the host galaxy of \sn in the context of host galaxies of SNe Ia-CSM from the ZTF. All hosts of SNe Ia-CSM are fairly typical of star-forming galaxies, as demonstrated by their location with respect to the main sequence of star-forming galaxies (grey-shaded band, \citealt{Elbaz2007a}). The host of \sn is an exception. It is among the more massive hosts, and the absence of measurable star-formation activity is uncommon of SN Ia-CSM host galaxies.}
    \label{fig:host_SFRvM}
\end{figure}

Figure~\ref{fig:host_SFRvM} shows the mass and SFR of \sn's host and those of host galaxies of the SNe Ia-CSM from the ZTF \citep{Kool2022, Sharma2023}. The host of \sn is not only one of the more massive hosts, the absence of measurable star-formation activity is also uncommon for SN Ia-CSM host galaxies. This result also stands out considering that the Type Ia-CSM-He SN 2020eyjh occurred in a star-forming dwarf galaxy. This difference was seen in the comparison of the colours and masses of the hosts by \citet[][their Fig. 10, lower-left panel, where \sn is the upper-right orange triangle and SN 2020eyjh is the lower-left triangle]{Dimitriadis2025}.
The lack of measureable star-formation activity poses additional constraints on the progenitor channel for \sn.

\section{Discussion} \label{sec:discussion}

\subsection{Late rebrightening}\label{sec:Late_rebrightening}

The existing sample of 28 SNe Ia-CSM \citep{Sharma2023, Silverman2013} all follow a similar photometric evolution that deviates significantly from the decline rates predicted for the radioactive decay of $^{56}$Ni \cite[see Fig.~5 of][]{Sharma2023}. All these objects have a slow evolution, sometimes even reaching a prolonged plateau phase (e.g. SN~2020eyj). 

\sn is the only Type Ia-CSM reported to exhibit a strong rebrightening. This unique characteristic is shown in Fig. \ref{fig:lightcurve_comparison}, where \sn is compared with the delayed SNe Ia-CSM PTF11kx and 2020eyj. All three objects deviate from the expected SN Ia decline rate after $\sim 25$~d post peak, when the CSM interaction begins dominating the light curves. The CSM interaction is more pronounced in the $g$ band (Fig.~\ref{fig:lightcurve_comparison}; upper panel), explaining the colour evolution toward the blue discussed in Sect.~\ref{sec:lightcurve}. The similarities between the other delayed SNe Ia-CSM  and \sn end there, as the latter is clearly rebrightening once the CSM interaction begins to dominate the powering.

We can get an estimate of the radiated energy during the two phases where the different powering mechanisms dominate. Integrating the pseudobolometric light curve from first detection until $\sim25$~d post peak gives $E_{\mathrm{rad, Ia}} \approx 4.6\times10^{49}$~erg for the radioactivity-dominated part of the evolution. When we integrate the second peak of the light curve ($\sim 25$~d post peak until the end of the evolution) we get $E_{\mathrm{rad, CSM}}\approx 1.1\times10^{50}$~erg. 
Typically, the kinetic energy in SNe Ia is on the order of $10^{51}$~erg \citep[][ and references therein]{Liu2023RAA}, which means that during the late part of the evolution, the mechanism that powers the light curve (CSM interaction) is efficient at converting kinetic energy into radiated energy.

As mentioned in Sect.~\ref{sec:latespectra}, the 
first CSM-dominated spectrum was obtained at +81 days, only after it was realised that \sn was not a normal SN Ia. 
The difficulty of catching rebrightening objects was also highlighted for CCSNe by \cite{sollerman2020}, and \sn was in fact noted already in the compilation by \cite{Soraisam2022}. 
\cite{Terwel2025} showed that we can discover SNe Ia-CSM emission by careful inspection and coaddition of late-time data.
They explicitly note \sn as the only SN Ia-CSM in their sample with a rebrightening light curve.
However, the need to react in real time to summon follow-up resources remains an issue for this kind of research. At the end of the day, the rebrightening of \sn was discovered semiserendipitously.

\subsection{Outside the optical regime}\label{sec:outside_optical}

Ultraviolet (UV) signatures are expected during shock/CSM interaction \citep[e.g.][]{graham2019a}. \sn was imaged with the {\it Hubble Space Telescope} in the UV with WFC3/UVIS F275W, 480 days after peak brightness (59693 MJD), to search for late-time interaction. Such interaction would suggest the existence of additional CSM at even larger radii. The observations were carried out as part of a snapshot proposal (1200~s; GO-16657, P.I. C. Fremling), but no object was detected. 

Mid-infrared (MIR) emission from the shock-heated dust (if present) has been interpreted as another proxy for CSM interaction. SN Ia-CSM 2018evt was detected in the MIR and proposed to show evidence for dust formation \citep{Wang2024}, and the compilation of \cite{Mo2024} showed that MIR detections by \textit{NEOWISE} are not uncommon among the nearby SNe Ia-CSM. The image-subtraction pipeline, developed by De et al. (in prep.), did not yield a \textit{NEOWISE} detection, neither during the period when \sn was optically visible, nor in the subsequent years after its optical fading.

Radio synchrotron emission arising at the shock front of ejecta/CSM interaction is another possible CSM-interaction signature. So far, the only SN Ia that has been detected at radio wavelengths is SN 2020eyj \citep{Kool2022}. \cite{Griffith2025arXiv250619071G} show that radio observations of a larger sample of SNe Ia-CSM did not yield additional detections. \sn was not included in the \cite{Griffith2025arXiv250619071G} sample owing to its large distance of almost 600 Mpc, and this is also what makes follow-up observations in other wavelength regimes insensitive for \sn.

\subsection{Missing H and He, nomenclature}\label{sec:nomenclature}

For stripped-envelope SNe, the sequence of stripping from SN Ib to SN Ic has an equivalence among the CSM-interacting transients, where the subclasses of H-depleted SNe Ibn and He-depleted SNe Icn now have been proposed to be joined by the even more stripped Type Ien SN 2021yfj \citep{Schulze2025}.

Analogously, the family of SNe-Ia-CSM was augmented with the H-poor but He-rich SN 2020eyj \citep{Kool2022}, and in this paper by the H- and He-poor \sn. This could warrant a discussion of nomenclature along the lines of SNe Ia-CSM-II/b/c or SNe Ia-CSM-H,He,C/O.
It is reminiscent of the discussion by \cite{Nomoto2018SSRv..214...67N}, who introduced the putative SN Ia-He CSM class based on the potential He-star donor progenitor system. We discuss potential progenitor systems for \sn below (Sect.~\ref{sect:progenitors}), but note here that given the complete lack of narrow H and He lines, the system could be considered the first SN Ia-CSM-C/O.

\subsection{Spectral characteristics}\label{sec:spectral_characteristics}

There are several things to note when it comes to the late-time spectra of \sn. The pseudocontinuum is exceptionally smooth compared to the SN Ibn models of \cite{Dessart2022A&A...658A.130D}, where the zigzag appearance of the iron forest reflects the low velocity of the cold dense shell (CDS). Furthermore, the \ion{Ca}{ii} NIR triplet in \sn is very broad, similar to other interacting SNe I (see Fig.~\ref{fig:csm_spectrum_comparison}). We note that the SN Ibn models by \citet[][see their footnote 2]{Dessart2022A&A...658A.130D} were unable to reproduce this feature at all.

We estimated the offset between the bluest line of the \ion{Ca}{ii} NIR triplet ($\lambda_{\mathrm{rest}} = 8498$~\AA) and the wavelength at zero intensity of the feature ($\lambda_{\mathrm{Z.I.}} \approx 8167 \pm 25$~\AA) to be $\sim11,689 \pm 868$~km~s$^{-1}$. This high velocity also matches the position of the \ion{Ca}{ii H\&K} line, as shown in Fig.~\ref{fig:csm_spectrum_comparison}.

Figure~\ref{fig:csm_spectrum_comparison} shows a comparison of 
\sn during the interaction phase with other interacting Type I SNe. Regarding the  \ion{Ca}{ii} NIR triplet, the comparison shows that SNe Ia-CSM have similarly large widths as seen for \sn, whereas the comparison Type Ibn and Icn SNe seem to have less broad triplets. When the broad \ion{Ca}{ii} doublet appears in SNe Ia-CSM (e.g. SNe~2020aekp and~2020eyj), the velocity offset is generally not as large as for the Ca NIR feature, contrary to the case for \sn. The \ion{O}{i} $\lambda$8446 could be present in those spectra and blended with the triplet, which would make the $\sim12,000$~km~s$^{-1}$ offset an overestimate.

The estimated high velocity for the calcium emission lines in \sn are more typical of ejecta than of a slower CDS, as envisioned in the models by \citet{Dessart2022A&A...658A.130D} mentioned above. Velocities on that order also for the iron emission lines would explain the very smooth pseudocontinuum. The calcium lines must come from parts of the ejecta that are not swept up and decelerated by the reverse shock which forms during the interaction. 
One explanation that makes parts of the ejecta visible is the case of the photosphere receding deeper than the reverse shock. In that case, we can detect photons emitted at the outermost part of the freely expanding ejecta. This scenario would also explain the blueshift of the features, since the ejecta on the opposite side would be obscured. \cite{Dessart2016} explored this scenario in the context of Type IIn SNe, where their ``model A" at late times has a flux contribution from the unshocked ejecta as the CDS gradually becomes less optically thick.
Alternatively, geometrical effects and deviations from spherical symmetry can lead to the simultaneous viewing of the CDS and the ejecta.

\cite{Chugai2004} had to add a Ca-rich, optically thick zone in the shocked ejecta of their models of the Type Ia-CSM SN 2002ic, in order to reproduce the broad Ca NIR feature. In that scenario, the broadening of the calcium lines is an effect of the fragmented nature of the CDS due to Rayleigh-Taylor instabilities. They showed that the strength and width of these features require a large area ratio of the emitting region compared to the thin-shell approximation. This mechanism can also result in a slight blueshift if scattering in the ejecta is taken into account. Detailed spectral modelling is not the focus of the present investigation, where we instead attempt to understand the late-time light curve.

\subsection{Light-curve modelling}\label{sec:modelling}

The first peak of \sn is analogous to that of several well-studied SNe Ia (see the \texttt{SNooPy} fits with SN Ia templates in Sect.~\ref{sec:lightcurve}). As such, it can be explained as the typical light curve powered by radioactive decay of 
nickel, outlined in the classical models by \cite{arnett1982}.

In order to understand the powering mechanism of the late-time rebrightening,
we turn our attention to the interaction between ejecta and CSM. We first note that explaining the $\sim50$~d rise of the second peak simply by photon diffusion through optically thick CSM would imply a very high CSM mass; this is shown following \cite{Chevalier2011}. Assuming steady, optically thick wind-like mass loss with $\rho (r) = D r^{-2}$, the time until breakout starts and the rise time are both given by the diffusion time, 
\begin{equation}
    t_\mathrm{d} = 2 \ \mathrm{d} \left( \frac{\kappa}{0.1 \ \mathrm{cm^2~g}^{-1}} \right) \left( \frac{D}{5 \times 10^{16} \ \mathrm{g~cm}^{-1}} \right).
\end{equation}
The fiducial value of $\kappa = 0.1 \ \mathrm{cm^2~g}^{-1}$ (corresponding to the opacity of H-free ionised matter) gives $D = 2.5 \times 10^{18} \ \mathrm{g~cm}^{-1}$.
Furthermore, since the shock-crossing time is $R_\mathrm{CSM} / v_\mathrm{sh} \approx t_\mathrm{d}$, the fiducial value of $v_\mathrm{sh} = 10^4 \ \mathrm{km~s}^{-1}$ gives $R_\mathrm{CSM} \approx 4 \times 10^{15} \ \mathrm{cm}$.
Therefore, $M_\mathrm{CSM}$ is estimated as $M_\mathrm{CSM} = 4 \pi D R_\mathrm{CSM} \approx 70$~\msun. We consider this option to be unrealistic given that the CSM is also devoid of both H and He.

Recently, \cite{Moriya2023} suggested that the long rise times often observed in SNe with CSM-interaction signatures could be attributed to the shock's dynamical timescale instead of the diffusion timescale. In particular, they considered explosions where the luminosity is dominated by the interaction of CSM having a power-law density structure ($\rho_{\mathrm{CSM}} \propto r^{-s}$) and ejecta having a broken power-law density structure ($\rho_{\mathrm{ej, inner}} \propto r^{-\delta}$ and $\rho_{\mathrm{ej, outer}} \propto r^{-n}$). Such a configuration leads to an asymptotic formula for the luminosity evolution in the form of Equation~\ref{eq:CSM_lum}, if one assumes that only the outer ejecta interact with the CSM and that a (constant) fraction of the kinetic energy is converted to radiation \citep[see][for a full derivation]{Moriya2023}:
\begin{equation}
    L = L_{1}t^{\alpha}
    \label{eq:CSM_lum}
\end{equation}
\noindent
Here, $\alpha = (6s-15+2n-ns)/(n-s)$ and $L_{1}$ is a constant depending on the parametrisation of the ejecta and CSM structures as well as the energy and mass in the explosion. \cite{Moriya2023} noted that for flat CSM configurations ($s<1.5$) and a range of fiducial values for $n$, $\alpha$ can take positive values. In such cases the luminosity would monotonically increase. Once the reverse shock enters the inner ejecta \citep[which typically has a flatter density profile, e.g.][]{Matzner1999}, $\alpha$ becomes negative and the luminosity declines. In this scenario, the rise time of the SN is dictated by the dynamical timescale of the shock. \cite{Chiba2024} generalised the formulation in \cite{Moriya2023} by including the effects of the extent of the CSM and diffusion. 

We used the \cite{Moriya2023} prescription to fit the second peak of \sn, assuming the effect of diffusion to be negligible. The parameters of the models were constrained to be consistent with the fact that here we are dealing with the explosion of a white dwarf (the model has previously mostly been used for Type IIn SNe). The parameters used for the theoretical light curve are as follows \citep[see][for definitions and details of the models]{Moriya2023}:
ejecta density structure with $n = 10$ and $\delta = 0$, $E_{\mathrm{kin}}=10^{51}$~erg, $M_{\mathrm{ej}}=1.3$~\msun, and kinetic energy conversion efficiency $\epsilon = 0.3$ \citep[e.g.][]{Fransson2014}. Given that the model assumes the interaction to dominate the luminosity, we initialised this powering mechanism at 30~d (Sect.~\ref{sec:Late_rebrightening}). This implies an evacuated region between the progenitor's surface and the CSM. With an ejecta velocity of 12,000~km~s$^{-1}$, this gives  $R_{\mathrm{CSM,~inner}}=3.1\times10^{15}$~cm. Similarly, assuming that the interaction ended  175--200~d after the explosion, based on the light-curve timescale, we have $R_{\mathrm{CSM,~outer}}=(9-10)\times10^{15}$~cm. 
\begin{figure}
\begin{center}
	    \includegraphics[scale=0.4]{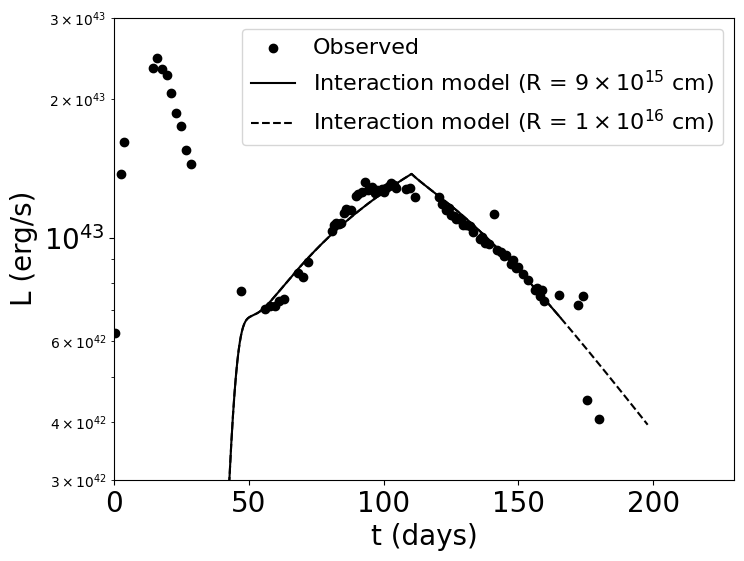} 
		\includegraphics[scale=0.4]{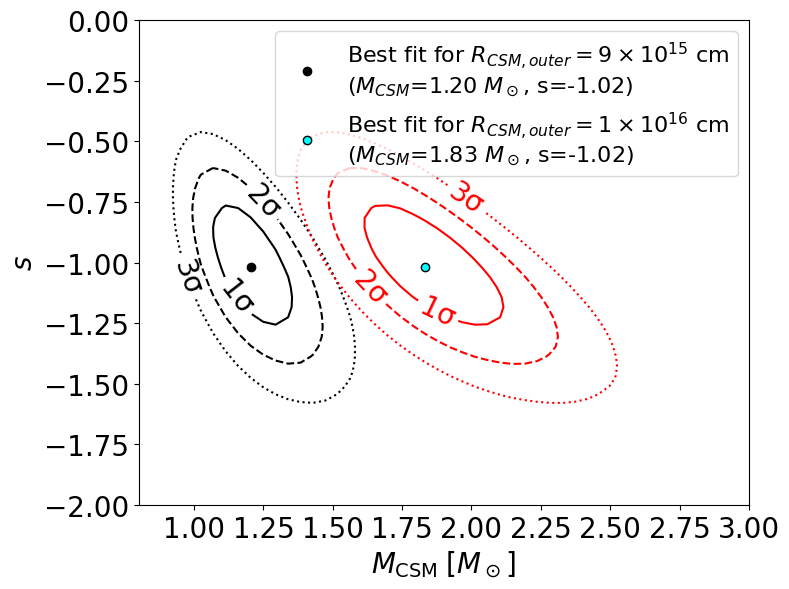}
\end{center}
	\caption{Best-fitted second peak of \sn (upper panel) with $R_{\mathrm{CSM,~inner}}=3.1\times10^{15}$~cm,  $R_{\mathrm{CSM,~outer}}=9\times10^{15}$~cm (solid), and $R_{\mathrm{CSM,~outer}}=1\times10^{16}$~cm (dashed). The choice of $R_{\mathrm{CSM,~outer}}$ is irrelevant to the shape of the light curve. The lower panel shows the confidence intervals that the $\chi^2$ fit yields for $R_{\mathrm{CSM,~outer}}=9\times10^{15}$~cm (black) and $R_{\mathrm{CSM,~outer}}=1\times10^{16}$~cm (red).}
	\label{fig:CSM_fits}
\end{figure}

We allowed the mass of the CSM, ($M_{\mathrm{CSM}}$) and the index of the CSM's density structure ($s$) to be the free parameters.  We solved numerically the momentum equation to compute the light curve \citep[see][for details]{Moriya2023} for each combination of the free parameters. The procedure was repeated for both $R_{\mathrm{CSM,~outer}}=9\times10^{15}$~cm and $R_{\mathrm{CSM,~outer}}=1\times10^{16}$~cm. 

The best $\chi^2$ light-curve fit is shown in Fig.~\ref{fig:CSM_fits} (upper panel).  Since the shape of the light curve is determined solely by the density profile, the choice of $R_{\mathrm{CSM,~outer}}$ is not affecting the light curve morphology. The inferred values are $M_{\mathrm{CSM}}=1.20$~\msun~for $R_{\mathrm{CSM,~outer}}=9\times10^{15}$~cm ($M_{\mathrm{CSM}}=1.86$~\msun~ for $R_{\mathrm{CSM,~outer}}=1\times10^{16}$~cm). The inferred value of the CSM density slope is $s = -1.02$, regardless of the choice for $R_{\mathrm{CSM,~outer}}$.
The total optical depth of CSM $\tau_\mathrm{CSM}$ when assuming $\kappa = 0.1$~cm$^{2}$~g$^{-1}$ is $\tau_\mathrm{CSM} = 0.42 \ (0.53)$, consistent with the assumption of neglecting the effect of diffusion.
The lower panel of Fig.~\ref{fig:CSM_fits} shows the corresponding 1--3$\sigma$ confidence intervals.

It is possible that the interaction began earlier but had a negligible contribution to the light curve. We repeated the modelling, initialising the interaction at $10$~d. With this assumption, the interaction terminating at $t \approx 175\, (-200)$~d now corresponds to larger values of $R_{\mathrm{CSM,~outer}}=10\,(-11)\times10^{15}$~cm. The obtained best-fit parameters for each value of $R_{\mathrm{CSM,~outer}}$ were $M_{\mathrm{CSM}}=1.38\,(2.31)$~\msun~, with $s = -2.23$ in both cases. In other words, if earlier onset of interaction is assumed, the requirement for larger $R_{\mathrm{CSM,~outer}}$ and steeper $s$ requires slightly larger values of $M_\mathrm{CSM}$. 

We see that at least qualitatively this interaction model can explain the timescale and luminosity of the second peak, even if no fine tuning has been done to fit the details. The amount of mass in the CSM to sustain this prolonged interaction is relatively modest at $\sim 1$--2~\msun. An important assumption in this model is spherical symmetry, which could be inaccurate in the case of \sn. If the CSM is asymmetrically distributed (e.g. confined in a disk), then we can expect a lower $M_\mathrm{CSM}$. The unusual and conspicuous feature of this interpretation is the negative slope of the required CSM density structure. For a canonical value of the ejecta slope ($n=10$), a steep enough rise to reproduce the luminosity evolution of the second peak requires a density that increases outward. We discuss these implications below.

\subsection{Progenitor scenarios}\label{sect:progenitors}

A key question for this system is what kind of information we can deduce for the progenitor given the long-lasting debate on the origin of Type Ia SNe, and in particular the nature of the binary system. There is a very extensive literature on this subject; here we briefly refer to \citet[][]{Liu2023RAA} and
\citet[][]{2025A&ARv..33....1R}
for recent reviews. The two suggested scenarios are a single-degenerate system (SD) where the exploding white dwarf (WD) accretes matter from a nondegenerate star through Roche-lobe overflow, or a double-degenerate (DD) scenario where two WDs merge to trigger the thermonuclear explosion. We briefly discuss both of these below.

\subsubsection{Single degenerate}

Perhaps the most natural suggestion for a SN Ia progenitor system is accretion from a normal star onto the WD \citep{Nomoto2018SSRv..214...67N}. A long-standing issue with this proposal is the lack of detections of interaction signatures anticipated for such systems. As the SN Ia explosion interacts with the CSM from the companion wind, the expected signatures in the early-time light curve \citep{Hayden2010ApJ...712..350H}, late-time spectra \citep{Lundqvist2015}, or at radio wavelengths \citep{Chomiuk2016} have not been seen.  As such, SNe Ia-CSM as a group is perhaps the best observational evidence that this channel exists \citep{Silverman2013,Sharma2023}, since the interpretation has generally been that the H-rich CSM must come from the wind of the donor star.
The radio detection of SN 2020eyj \citep{Kool2022} is the best evidence to date for a nondegenerate He-star donor to a SN Ia. The clear spectroscopic He signatures in conjunction with the late-time CSM-driven light curve and the radio detection all point to significant amounts of helium-rich CSM as envisioned in such a single-degenerate system \citep{Nomoto2018SSRv..214...67N}.

The dramatic CSM-driven rebrightening of \sn could perhaps also be explained in such a scenario.  However, the significant difference is the complete lack of H and He, with the implications that the CSM must come from a completely stripped progenitor, reminiscent of the WR stars suggested as progenitors for CC SNe Ic or Icn. To the best of our knowledge, no such suggested system exists in the literature for a SN Ia progenitor.  The other unusual aspect for \sn is the fact that our modelling implies a density structure of the CSM that is not expected for a steady wind from any companion. To obtain a CSM that has an increasing density structure out to some radius requires a mass-loss mechanism that was more efficient in the past. To create the required CSM would requires a mass-loss rate that effectively decreased by a factor of 1000 over the few years prior to the explosion, when the CSM was produced. This calls for a more eruptive outflow, which also echoes the need for an earlier nonstandard mass-loss mechanism to generate the H- and He-free progenitor in the first place. 

An alternative scenario invokes inspiralling of the companion WD inside the envelope of an asymptotic giant branch (AGB) star, during a common-envelope (CE) phase \citep{Paczynski1976}. \cite{Kashi2011} suggested that some amount of the CE will not be ejected and fall back. The bound material will interact with the core-WD binary, and reduce its orbital separation on shorter timescales than the dynamical post-CE phase. This can lead to a merger and a SN Ia during the planetary-nebula phase, a progenitor system coined core-degenerate \citep[CD;][]{Kashi2011}. If the WD penetrates deep within the carbon layer and ejects it, it might explain the uncommon composition of the CSM of \sn. Still, the absence of H and He signatures would require an extensive stripping of the outer layers prior to the CE phase.

Regardless of the mechanism that leads to a thermonuclear explosion, the amount of $M_{\rm CSM}$ points to a heavy nondegenerate companion. A star on the heavier end of intermediate-mass stars must have formed recently. The upper limit on the SFR estimated for the host (Sect.~\ref{sec:host}) seems to disfavour recent star formation. However the large distance to the host prevents more precise estimates of the SFR at the location of the explosion.

\subsubsection{Double degenerate}\label{sec:DD}

Whereas strong CSM interaction would typically be interpreted as evidence for a nondegenerate binary, there are reasons to mention also the DD scenario for \sn. The CSM is apparently completely devoid of H and He, which is a highly unusual composition. The potential presence of C/O in the CSM makes the scenario of a companion WD an intriguing one. Moreover, the light-curve modelling (see Sect.~\ref{sec:modelling}) shows that the amount of material required could be less than the Chandrasekhar mass, $M_{\mathrm{Ch}}$, which does not immediately rule out a DD scenario. However, a $M_{\mathrm{CSM}}=1.20$~\msun\ 
scenario would basically require the complete disruption of a near-Chandrasekhar-mass companion WD prior to the explosion. The negative index density structure also points to a catastrophic mass-loss episode rather than a steady wind.

A C/O-rich CSM formed during the merger of two WDs has been invoked in the past to explain SN 2003fg-like SNe Ia \citep[e.g.][]{Dutta2022, Dimitriadis2022, Siebert2024}. Those events, also coined super-Chandrasekhar, show absorption features of unburned carbon in their early spectra. Furthermore, some of the light curves show an early flux excess \citep{Jiang2021,Dimitriadis2023, Srivastav2023}, similar only to their related SN 2002es-like SNe \citep{Hoogendam2024}. While a C/O CSM could explain SN 2003fg-like events \citep[e.g.][]{Noebauer2016}, neither the distance nor the amount of CSM is similar to that required for \sn.

A gravitational-wave-driven merger between two WDs is a relatively gentle process. During this inspiralling, the secondary WD might expel some material through tidal tail ejection \cite{RaskinKasen2013}. If the material is relatively nearby ($\lesssim 10^{14}-10^{16}$~cm), interaction with the ejecta could have observational signatures. The amount of CSM expected from such an interaction is hardly larger than 0.01 \msun, which appears too low to explain the late-time light curve of \sn. Additionally, the unbound material in the tidal tail should have WD escape velocities which are $\sim5000$~km~s$^{-1}$ while the FWHM of the narrow lines in \sn is $\lesssim1500$~km~s$^{-1}$.

Dynamical encounters are more violent. These may occur in locations of large stellar densities such as
globular clusters or galactic nuclei \citep[see][]{lyndenbell67,meylan97,alexander05}. \cite{Rosswog2009ApJ...705L.128R} explored the scenario of a near-central collision ($R_{\rm peri} \lesssim (R_1 + R_2)/3$) of two WDs that could occur in such an environment; here,
$R_1$ and $R_2$ are the radii of the WDs and $R_{\rm peri}$ is the pericentre of the stellar centre of mass.
They find that the light curves modelled from such collision-induced thermonuclear explosions can resemble observed SN Ia light curves. 

However, encounters with larger initial pericentre distances are more common. If such a passage is close enough to raise tides on the stars, they become tidally spun up at the expense of orbital energy. With each passage the orbit becomes tighter, the stars spin up, and  they eject increasingly larger amounts of mass. Eventually, the lighter star is disrupted and forms a disk -- or, if the final impact is central enough, it triggers a version of the above described collision-induced thermonuclear explosion. 

\begin{figure}
\centering
    \includegraphics[width=\hsize]{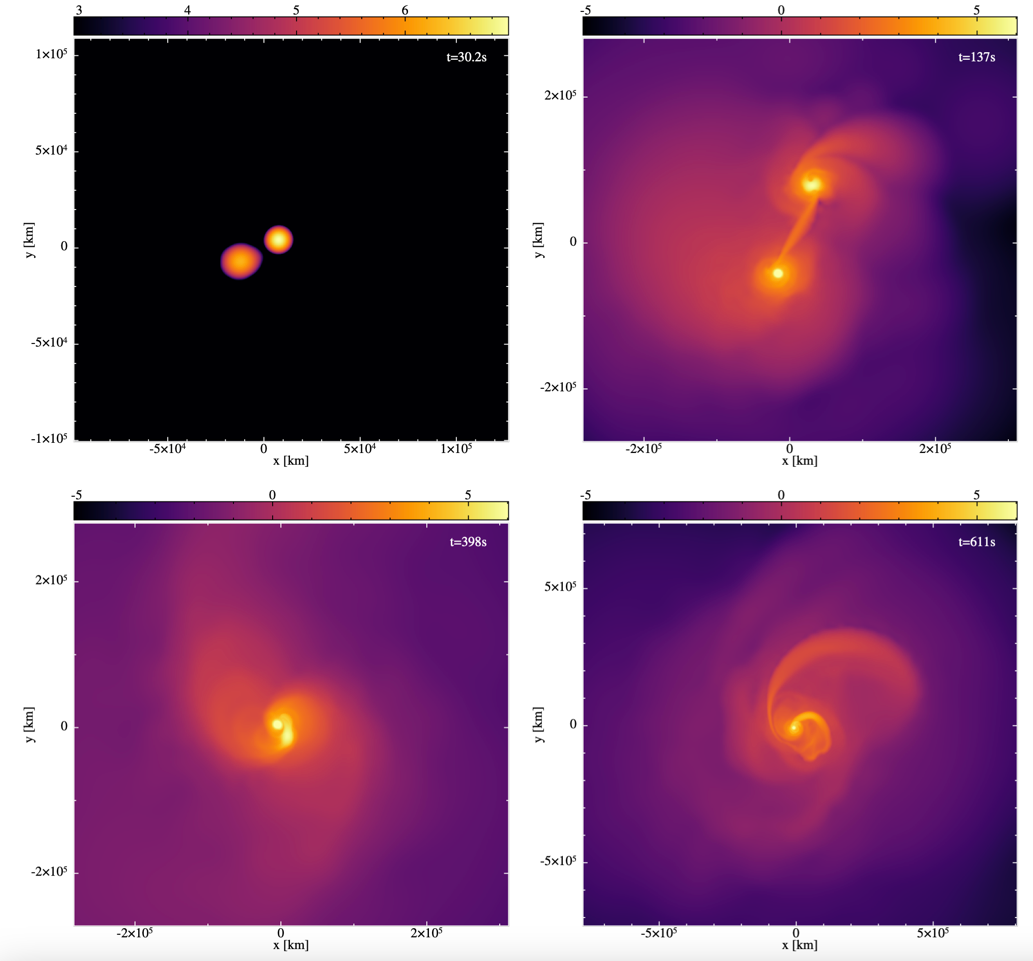}
    \caption{Density maps of a sequence of fly-by encounters of a 0.5~\msun\ and a 0.8~\msun\ WD. The initial pericentre distance was $R_{\rm peri} = R_1 + R_2$. The simulation was carried out with the \texttt{MAGMA2} code \citep{rosswog20}.}
    \label{fig:fly_by}
\end{figure}

The secondary WD would provide (nearly exclusively) the amount of CSM required for SN 2020aeuh in such an encounter, so it would have to be above $\sim 0.5$~\msun.
As an illustration of the overall dynamics of a parabolic fly-by, we show in Fig.~\ref{fig:fly_by} the grazing collision
($R_{\rm peri} = R_1 + R_2$) of two CO-WDs with 0.5 and 0.8~\msun\ on an initially parabolic orbit. The
simulation is performed with the \texttt{MAGMA2} code \citep{rosswog20}. In this illustrative example, the lighter WD suffers
several close encounters, each time shedding a substantial amount of mass into space.

Even in such encounters, it is hard to see how $>1~$\msun\ can be produced or how the CSM would be detached from the progenitor to lead to a delayed SN Ia-CSM. However, this example serves as a proof of concept that such fly-bys can comfortably produce a CO-CSM of a few tenths of solar mass. A more extensive exploration of the parameter space (masses, mass ratios, impact parameters) is outside the scope of this work.

\section{Summary and conclusions} \label{sec:summary}

We have reported observations of \sn, which is a Type Ia-CSM with spectacular properties. After an initial phase where both the spectrum and the light curve reveal a thermonuclear SN Ia, the light curve rebrightens and a 100+ days long CSM-driven phase is entered. The two phases are well separated, providing evidence for a detatched dense shell of CSM surrounding the SN Ia explosion. Unlike any previous SN Ia-CSM, neither hydrogen nor helium features can be seen in the high-quality spectra of \sn during the interaction-dominated phase. 

During the early phase of its evolution, \sn appeared as a slightly overluminous SN Ia (Sect.~\ref{sec:lightcurve}). Its photometric properties during the first peak placed it marginally within the width-luminosity relationship of normal SNe Ia \citep{phillips1993}, near the outlying subtypes of SN 2003fg-like, SN 1991T-like, and Ia-CSM SNe \citep[see Fig.~1 of][]{Taubenberger2017}. While its later evolution revealed it as a SN Ia-CSM, its classification spectrum is best matched to SN 1991T-like SNe. Furthermore, the H- and He-poor CSM is reminiscent of the postulated configuration for the progenitors of SN 2003fg-like SNe, albeit with vastly different properties (Sect.~\ref{sec:DD}). \sn exploded in the outskirts of its host galaxy, like some SN 2003fg-like events \citep{Khan2011}. In contrast to other SNe Ia-CSM, the host of \sn is an early-type galaxy with a very low SFR (Sect.~\ref{sec:host}).

The rebrightening of \sn together with its spectral change uncovered its nature as a delayed SN Ia-CSM. Its unique late spectral appearance shows a few narrow lines, possibly of oxygen and carbon, and no evidence of hydrogen or helium. This makes \sn the first known Type Ia-CSM-C/O (Sect.~\ref{sec:nomenclature}). The estimated FWHM of the narrow lines is on the order of $10^{3}$~km~s$^{-1}$ (Sect.~\ref{sec:latespectra}), consistent with the narrow component of H$\alpha$ in the \cite{Sharma2023} sample of SNe Ia-CSM. The nature of the broad calcium lines in the spectra of interacting SNe remains puzzling. Their estimated velocities in \sn are typical for fast-moving ejecta. Seeing the ejecta and the CDS simultaneously can be an effect of geometry or physical conditions of the CDS (Sect.~\ref{sec:spectral_characteristics}). 

To understand the powering mechanism responsible for the second peak, we fitted the pseudobolometric light curve with the CSM-interaction model from \cite{Moriya2023}. In this formalism, the timescale of the evolution is dictated by the dynamical timescale of the shock instead of diffusion. In the case of \sn, diffusion would require $\sim~70$~\msun\ of CSM, which we regard as unfeasible for an H- and He-free composition. The best-fitted \cite{Moriya2023} model instead yields 
 $M_{\mathrm{CSM}} \approx 1$--2~\msun~ (under a spherical symmetry assumption), with an unusual CSM structure with $s\lesssim-1$ (Sect.~\ref{sec:modelling}). While such a density structure for the CSM is uncommon, the model follows adequately the light-curve morphology (Fig.~\ref{fig:CSM_fits}). The amount of required CSM mass is in the regime where neither the SD nor the DD progenitor scenario can be excluded.

\sn is a unique SN Ia-CSM that does not quite fit the picture of any suggested progenitor system for thermonuclear explosions (Sect.~\ref{sect:progenitors}). The composition of the CSM could point toward a DD system, but the inferred masses call for a complete obliteration of a massive WD companion. Within the DD scenario, the case of multiple dynamical encounters can shed substantial amounts of material from the secondary WD (Sect.~\ref{sec:DD}). However, both the $M_{\rm CSM}$ we infer and the detached nature of the CSM are difficult to explain in this scenario. Perhaps a lower $M_{\rm CSM}$ could be required in the case of asymmetric CSM. Additionally, the delay time between the stripping of the companion WD (either in the tidal-tail ejection or the fly-by encounter paradigms) and the explosion might account for the large radius of the CSM. Either way, a systematic study of the mechanisms that could produce some CSM in the DD scenario is still lacking.  Regarding the SD (or CD) scenario, extensive stripping of the companion's H and He layers is required, and detailed stellar-evolution scenarios leading up to such a configuration are hitherto missing \cite[e.g.][]{Liu2023RAA}. The remarkably low SFR of the host (Sect.~\ref{sec:host}) further weakens the nondegenerate cases. Altogether, the exotic appearance and properties of \sn pose crucial difficulties to the progenitor pathways of SNe Ia-CSM in particular and of SNe Ia in general.

\section*{Data Availability}

All spectra are available at the WISeREP, via \href{https://www.wiserep.org/object/17853}{https://www.wiserep.org/object/17853}. The entire photometry table (Table~\ref{tab:photometry}) is available in electronic form at the CDS via anonymous ftp to cdsarc.u-strasbg.fr (130.79.128.5) or via \href{https://cdsarc.cds.unistra.fr/viz-bin/cat/J/A+A/704/A135}{https://cdsarc.cds.unistra.fr/viz-bin/cat/J/A+A/704/A135}.
\begin{acknowledgements}
We thank N. Sarin, A. Singh, S. Yang, B. van Baal, A. Gkini, S. Barmentloo, Y. Hu, and the Stockholm University Supernova Group for valuable feedback and discussions.

This work has been enabled by support from the research project grant ``Understanding the Dynamic Universe'' funded by the Knut and Alice Wallenberg under Dnr KAW 2018.0067.
E.C.K. acknowledges support from the G.R.E.A.T. research environment funded by {\em Vetenskapsr\aa det}, the Swedish Research Council, under project 2016-06012, and support from The Wenner-Gren Foundations.
S.R. has been supported by the Swedish Research Council (VR) under grant number 2020-05044 and by the research
environment grant ``Gravitational Radiation and Electromagnetic Astrophysical Transients'' (GREAT) funded by the Swedish Research Council (VR) under Dnr 2016-06012, by the Knut and Alice Wallenberg Foundation under grant Dnr. KAW 2019.0112, by Deutsche Forschungsgemeinschaft (DFG, German Research Foundation) under Gemany’s Excellence Strategy -- EXC 2121 ``Quantum Universe'' -- 390833306 and by the European Research Council (ERC) Advanced Grant INSPIRATION under the European Union’s Horizon 2020 research and innovation programme (grant agreement No.
101053985). 
A.V.F.’s group at U.C. Berkeley received financial assistance from the Christopher R. Redlich Fund, as well as donations from Gary and Cynthia Bengier, Clark and Sharon Winslow, William Draper, Timothy and Melissa Draper, Briggs and Kathleen Wood, Sanford Robertson (T.G.B. is a Draper-Wood-Robertson Specialist in Astronomy, Y.Y. was a Bengier-Winslow-Robertson Fellow in Astronomy), and numerous other donors.
Y.Y.'s research is partially supported by the Tsinghua University Dushi Program.

Based in part on observations obtained with the 48-inch Samuel Oschin Telescope and the 60-inch Telescope at the Palomar Observatory as part of the Zwicky Transient Facility project. ZTF is supported by the U.S. National Science Foundation (NSF) under grant  AST-2034437 and a collaboration including Caltech, IPAC, the Weizmann Institute for Science, the Oskar Klein Center at Stockholm University, the University of Maryland, Deutsches Elektronen-Synchrotron and Humboldt University, the TANGO Consortium of Taiwan, the University of Wisconsin at Milwaukee, Trinity College Dublin, Lawrence Livermore National Laboratories, and IN2P3, France. Operations are conducted by COO, IPAC, and UW.
The SED Machine at Palomar Observatory is based upon work supported by the NSF under grant 1106171.
We made use of the {\tt Fritz} platform \citep{VanderWalt2019, Coughlin2023ApJS..267...31C}.
The ZTF forced-photometry service was funded under the Heising-Simons Foundation grant \#12540303 (PI M. Graham). 

This work is based in part on observations made with the Nordic Optical Telescope, operated by the Nordic Optical Telescope Scientific Association at the Observatorio del Roque de los Muchachos, La Palma, Spain, of the Instituto de Astrofisica de Canarias.
The data presented here were obtained in part with ALFOSC, which is provided by the Instituto de Astrofisica de Andalucia (IAA-CSIC) under a joint agreement with the University of Copenhagen and NOTSA.

Some of the data presented herein were obtained at the W. M. Keck Observatory, which is operated as a scientific partnership among the California Institute of Technology, the University of California, and NASA; the observatory was made possible by the generous financial support of the W. M. Keck Foundation. We thank WeiKang Zheng for assistance with the Keck observations.

This work has made use of data from the Asteroid Terrestrial-impact Last Alert System (ATLAS) project. The Asteroid Terrestrial-impact Last Alert System (ATLAS) project is primarily funded to search for near-Earth objects through NASA grants NN12AR55G, 80NSSC18K0284, and 80NSSC18K1575; byproducts of the NEO search include images and catalogs from the survey area. This work was partially funded by Kepler/K2 grant J1944/80NSSC19K0112 and HST GO-15889, and STFC grants ST/T000198/1 and ST/S006109/1. The ATLAS science products have been made possible through the contributions of the University of Hawaii Institute for Astronomy, the Queen’s University Belfast, the Space Telescope Science Institute, the South African Astronomical Observatory, and The Millennium Institute of Astrophysics (MAS), Chile.
\end{acknowledgements}
\bibliography{main}

\begin{appendix}
\section{Photometry and spectroscopy}

\begin{table*}
\caption{\label{tab:photometry}Optical photometry of \sn, in observed magnitudes. Phase is relative to ZTF first detection, in rest-frame days.}
\centering
\begin{tabular}{ccccccc} \hline
        MJD &   Phase & Filter &     Magnitude & Error & Limiting magnitude & Telescope+Instrument \\ \hline
59195.5128 & -3.46 & r &       &      & 20.72 & P48+ZTF \\
59195.5328 & -3.45 & g &       &      & 20.91 & P48+ZTF \\
59198.4634 & -0.84 & r &       &      & 20.59 & P48+ZTF \\
59199.415  & 0     & r & 20.86 & 0.34 & 20.34 & P48+ZTF \\
59200.012  & 0.53  & g & 20.72 & 0.17 & 20.96 & P48+ZTF \\
59200.4896 & 0.95  & i & 20.64 & 0.74 & 19.27 & P48+ZTF \\
59200.5297 & 0.99  & r & 20.52 & 0.3  & 20.12 & P48+ZTF \\
59202.5143 & 2.75  & g & 19.75 & 0.07 & 20.87 & P48+ZTF \\
59203.016  & 3.2   & r & 19.88 & 0.09 & 20.85 & P48+ZTF \\
59203.5308 & 3.65  & g & 19.58 & 0.07 & 20.81 & P48+ZTF \\
59204.5268 & 4.54  & i & 19.76 & 0.11 & 20.44 & P48+ZTF \\
59206.5371 & 6.32  & g & 19.57 & 0.37 & 18.92 & P48+ZTF \\
59206.5979 & 6.38  & c & 19.57 & 0.08 & 20.57 & ATLAS   \\
59208.6259 & 8.18  & c & 19.28 & 0.05 & 20.9  & ATLAS   \\
59212.6063 & 11.71 & o & 19.14 & 0.11 & 19.82 & ATLAS   \\
59214.5459 & 13.43 & o & 19.32 & 0.15 & 19.69 & ATLAS   \\
59215.4762 & 14.26 & r & 19.2  & 0.12 & 19.77 & P48+ZTF \\
59215.4908 & 14.27 & g & 19.29 & 0.14 & 19.7  & P48+ZTF \\
59216.494  & 15.16 & i & 19.54 & 0.17 & 19.77 & P48+ZTF\\
\hline
\multicolumn{7}{c}{ } \\
\multicolumn{7}{c}{(This table is available at the CDS in its entirety.)}
\end{tabular}
\end{table*}

\begin{table*}
\caption{\label{tab:spec}Log of spectroscopic observations of \sn.}
\centering
\begin{tabular}{cccc} \hline
MJD     & Observation Date  & Phase past discovery   & Telescope + Instrument \\
        &                   &(rest-frame days)       & \\ \hline
59227.4 & 2021 Jan. 13       &        24.8          & OSMOS (from TNS)\\
59291.9 & 2021 Mar. 18       &        82.1        & NOT+ALFOSC  \\
59310.1 & 2021 Apr. 6        &       98.3        & NOT+ALFOSC  \\
59311.6 & 2021 Apr. 7        &       99.6        & Keck1+LRIS  \\
59344.5 & 2021 May 10       &       128.8        & Keck1+LRIS  \\
59346.0 & 2021 May 12       &       130.1        & NOT+ALFOSC  \\
59365.9 & 2021 May 31       &       147.8        & NOT+ALFOSC  \\
59404.3 & 2021 July 9        &       181.9        & Keck1+LRIS \\
\hline
\end{tabular}
\end{table*}

\begin{table*}
\caption{{\it Left:} Host-galaxy photometry. Magnitudes in the AB system and not corrected for reddening.\\
{\it Right:} Inferred host parameters from SED fitting with \texttt{Prospector}.}\label{tab:host}

\centering
\begin{tabular}{cc|cc} \hline
Filter     & Magnitude & Parameter & Best-fitted value\\ [5pt]
 \hline
SDSS/$u$  &$ 20.07 \pm 0.21 $ & Age~[Myr] & $7808^{+3725}_{-3636}$\\[3pt]
SDSS/$g$  &$ 19.10 \pm 0.06 $& $\log(M)~$[\msun] & $10.50^{+0.13}_{-0.18}$\\[3pt]
SDSS/$r$  &$ 18.05 \pm 0.03 $& SFR~[\msun~yr$^{-1}$] & $0.00^{+0.12}_{-0.00}$\\[3pt]
SDSS/$i$  &$ 17.72 \pm 0.04 $& $E(B-V)_{\mathrm{star}}$ [mag] & $0.09^{+0.09}_{-0.06}$\\[3pt]
SDSS/$z$  &$ 17.49 \pm 0.18 $& & \\[3pt]
WISE/$W1$ &$ 17.74 \pm 0.04 $& & \\[3pt]
WISE/$W2$ &$ 18.25 \pm 0.08 $& & \\[3pt]
\hline
\end{tabular}
\end{table*}

\end{appendix}

\end{document}